\documentclass[fleqn,usenatbib]{mnras}

\usepackage{newtxtext,newtxmath}

\usepackage[T1]{fontenc}

\DeclareRobustCommand{\VAN}[3]{#2}
\let\VANthebibliography\thebibliography
\def\thebibliography{\DeclareRobustCommand{\VAN}[3]{##3}\VANthebibliography}

\usepackage{graphicx}	
\usepackage{amsmath}	
\usepackage{subcaption}
\usepackage{hyperref}

\title[$f(R)$ Gravity with Future 21 cm]{Constraining Scale-Dependent Growth in $f(R)$ Gravity with Future 21 cm Surveys}

\author[et al.]{Apurba Samanta
,$^{1}$\thanks{E-mail: apurbasamanta79@gmail.com }
Bhuwan Joshi,$^{2}$ \thanks{E-mail: bhuwanj230@gmail.com}
Jess Worsley,$^{3}$\thanks{E-mail: wrsjes002@myuct.ac.za}
Peter Dunsby $^{3,6,7}$\thanks{E-mail: peter.dunsby@uct.ac.za}
and Saikat Chakraborty$^{4,5,6}$ \thanks{E-mail: saikat.chakraborty@nwu.ac.za}
\\
$^{1,2}$ School of Physical Sciences, Indian Institute of Technology Mandi,  Himachal Pradesh, 175005, India\\
$^{3}$ Department of Mathematics and Applied Mathematics, University of Cape Town, Rondebosch 7700, Cape Town, South Africa\\
$^{4}$ Institute of Research and Development, Duy Tan University, Da Nang 550000, Vietnam\\
$^{5}$ Faculty of Natural Sciences, Duy Tan University, Da Nang 550000, Vietnam\\
$^{6}$ Center for Space Research, North-West University, Potchefstroom 2520, South Africa \\
$^{7}$ South African Astronomical Observatory, Observatory 7925, Cape Town, South Africa
}

\date{Accepted XXX. Received YYY; in original form ZZZ}

\pubyear{\the\year{}}

\begin{document}
\label{firstpage}
\pagerange{\pageref{firstpage}--\pageref{lastpage}}
\maketitle

\begin{abstract}
Recent observations, particularly from DESI, have provided intriguing hints of dynamical behaviour in late-time dark energy. Modified gravity theories offer a compelling framework for interpreting such phenomena, with $f(R)$ gravity emerging as one of the most extensively studied examples. A central challenge in these models, however, lies in determining the precise functional form of $f(R)$. Nevertheless, several viable models have been proposed that successfully reproduce the standard $\Lambda$CDM cosmology at high red shifts while generating late-time cosmic acceleration without an explicit dark energy component. Within this framework, the evolution of the linear matter density contrast becomes scale dependent, leading to a growth index that varies with both scale and redshift. In this work, we explore the capability of forthcoming 21\,cm observations to constrain the growth index, as well as the combined neutral hydrogen (HI) bias and growth-rate parameter. Our results indicate that future 21\,cm surveys can provide meaningful, though moderate, support for these modified gravity scenarios.

\end{abstract}




\section{Introduction}
\label{sec:intro}

Einstein’s theory of general relativity (GR) has long served as the cornerstone of modern gravitational physics. Within this framework, the $\Lambda$ cold dark matter ($\Lambda$CDM) model successfully describes a broad range of cosmological observations. According to the standard paradigm, ordinary baryonic matter contributes only $\sim 4\%$ to the total energy density of the Universe, whereas dark matter and dark energy account for approximately $20\%$ and $76\%$, respectively. Despite its phenomenological success, the physical nature of the dark sector remains elusive, and the $\Lambda$CDM model is plagued by several fundamental theoretical shortcomings. Early studies by \citep{Dicke1961} identified a severe fine-tuning issue, demonstrating that even a marginally larger value of $\Lambda$ would have inhibited the formation of cosmic large-scale structures. The model also suffers from the `coincidence problem'—the unexplained near-equality of the present-day matter and dark energy densities, despite their markedly different evolutionary behaviours throughout cosmic history \citep{Steinhardt1998}. Furthermore, $\Lambda$CDM faces the vacuum energy problem, which stems from the enormous mismatch between the observed vacuum energy density and the zero-point energy predicted by quantum field theory \citep{Adler1995}. More recently, early observations from the Dark Energy Spectroscopic Instrument (DESI) have provided tentative indications of dynamical dark energy, favouring a time-dependent equation of state within a $w_0 w_a$CDM parametrisation \citep{Carloni2024, Luongo2024, Tada2024}. Motivated by these theoretical and observational tensions, there is considerable interest in modified gravity (MG) theories, where late-time cosmic acceleration arises not from an exotic dark energy fluid, but from modifications to the gravitational sector itself. Historically, attempts to construct a more complete and quantum-consistent theory of gravity motivated early extensions by Weyl and Eddington \citep{eddington1923}, while later work by \cite{PhysRevD.16.953} demonstrated that the inclusion of higher-order curvature invariants renders the gravitational action renormalisable. Among this wide class of extended theories, $f(R)$ gravity \citep{10.1093/mnras/150.1.1} has emerged as one of the most extensively studied frameworks. This class of theories underpins several highly influential cosmological models, ranging from Starobinsky inflation in the early Universe \citep{Starobinsky:1980te} to the Hu--Sawicki model \citep{PhysRevD.76.064004} proposed to explain late-time acceleration. In the context of $f(R)$ gravity, the modified field equations can be recast into an Einstein-like form containing an `effective curvature fluid' \citep{PhysRevD.70.043528}. Such a fluid can violate the strong energy condition, naturally driving an accelerated expansion phase without the explicit need for a cosmological constant. However, despite their theoretical appeal, $f(R)$ theories yield fourth-order differential field equations that are notoriously difficult to solve analytically. To circumvent these challenges, dynamical systems techniques \citep{Carloni_2005,Carloni:2006mr} provide a useful framework for studying specific subclasses of solutions, though their applicability is often limited to particular functional forms of $f(R)$. Alternatively, cosmographic approaches \citep{article} have been developed to characterise the expansion history using directly observable kinematic quantities (such as the Hubble, deceleration, jerk, and snap parameters) without initially specifying the gravitational action, though closing the system still requires assuming a background expansion history. Within this landscape, $\Lambda$CDM-mimicking $f(R)$ models are of particular phenomenological interest. Motivated by the recent work of \cite{MacDevette:2024wpg} — who derived the evolution equations for the matter density contrast in these mimicking models using the 1+3 covariant formalism—we investigate the distinct observational signatures of this theoretical framework. A fundamental prediction of such MG models is that the evolution of the matter density contrast becomes explicitly scale-dependent. This contrasts starkly with the standard $\Lambda$CDM paradigm, wherein the linear growth index is scale-independent and remains approximately constant ($\gamma \simeq 0.55$). Instead, these $f(R)$ models yield a generalised linear growth index, $\gamma(k,z)$, and a corresponding growth-rate parameter, $\beta(k,z)$, that exhibit explicit dependence on both redshift, $z$, and co-moving wave number, $k$. 

This growth index has been studied previously \citep{Dossett:2010gq,Yin:2019rgm} and favours a variation in $\gamma$. Recently, the \citep{Euclid:2025bxg} analysis from the Euclid survey also tested $\gamma$ within the framework of modified gravity theories. In this work, we use the 21cm signal from neutral hydrogen as a new window to probe the Universe, spanning epochs from the cosmic dark ages to the post-reionisation era \citep{Pritchard_2012,21cm_BOOK,21cm_review,10.1093/mnras/227.1.1}. Here we build upon this foundation by quantifying the capability of forthcoming 21-cm intensity mapping (IM) surveys, such as those planned for the Square Kilometre Array (SKA), to constrain these scale-dependent growth signatures. Because 21-cm IM efficiently traces the underlying large-scale structure over broad redshift windows without the severe shot-noise limitations of traditional discrete galaxy surveys, it is exceptionally sensitive to redshift-space distortions (RSD) \citep{10.1093/mnras/227.1.1}. By exploiting the sensitivity of the RSD signal to $\gamma(k,z)$, we assess the extent to which these $f(R)$ models can be distinguished from standard cosmology, thereby establishing a robust, independent avenue for testing deviations from GR.

The remainder of this paper is structured as follows. In Section~\ref{sec:F(R)gravity}, we review the theoretical framework of $\Lambda$CDM-mimicking $f(R)$ gravity and outline the corresponding evolution equations for the matter density contrast and fixing the initial condition Section~\ref{sec:21cm_Basics} presents the fundamentals of 21-cm cosmology along with the modelling of the statistical power spectrum. Our forecasting methodology, based on the Fisher matrix formalism, is detailed in Section~\ref{sec:results_and_discussion} alongside our primary results and discussion. Finally, we summarise our main conclusions in Section~\ref{sec:conclusion}.

\section{$\Lambda$CDM expansion history in $f(R)$ gravity}\label{sec:F(R)gravity}

Following \cite{Chakraborty:2021jku,MacDevette:2024wpg,Chakraborty:2025lkz}, we give a very brief review of $f(R)$ gravity with an expansion history identical to $\Lambda$CDM. Our goal here is to present the reader with the minimal set of equations that will be used later on in the subsequent 21-cm cosmological analysis. For a more detailed exposition, the reader is referred to the references \citep{Chakraborty:2021jku,MacDevette:2024wpg,Chakraborty:2025lkz}. In the late time context where the effective curvature fluid serves as a dark energy component, it is conceptually clearer to express the cosmological field equations of $f(R)$ in the following form
\begin{subequations}
    \begin{eqnarray}\label{field_eqs_new}
        3H^{2} = \rho_{\rm tot} = \rho + \rho_{\rm DE}\,,\label{eq:fried_new}\\
        -\left(2\dot{H} + 3H^{2}\right) = P_{\rm tot} = P_{\rm DE}\,,\label{eq:Raychoudhuri_new}
    \end{eqnarray}     
\end{subequations}
where $\rho$ is the energy density of the non-relativistic matter, and we have defined the energy density and pressure of the dark matter as follows
\begin{subequations}
    \begin{align}
         \rho_{\rm DE}&=\frac{1}{2}(Rf'-f)-3H\dot{F}+3H^{2}(1-f')\,,\label{eq:DEed}\\
         P_{\rm DE}&=\ddot{f'}+2H\dot{f'}-\frac{1}{2}(RF-f)-(2\dot{H}+3H^2)(1-f')\,. \label{eq:DEp}
    \end{align}
\end{subequations}
The dark energy equation of state $w_{\rm DE} = \frac{P_{\rm DE}}{\rho_{\rm DE}} = - \frac{2\dot{H}+3H^{2}}{3H^{2}-\rho}$ can be expressed in two alternative ways:
\begin{subequations}
\begin{align}
    w_{\rm DE} &= \frac{\ddot{f'}+2H\dot{f'}-\frac{1}{2}(Rf'-f)-(2\dot{H}+3H^2)(1-f')}{\frac{1}{2}(Rf'-f)-3H\dot{f'}+3H^{2}(1-f')}
    \\
    &= \frac{2q-1}{3-3\Omega_m}\,,
\end{align}
\end{subequations}
where $\Omega_m=\frac{\kappa\rho}{3H^2}$ and $q=-\frac{\dot{H}}{H^2}$ is the cosmographic deceleration parameter. The next order cosmographic parameter is the jerk parameter $j$:
\begin{equation}
    j = \frac{\ddot H}{H^3} - 3q - 2 = 2q^2 + q + (1+z)\frac{dq}{dz}\,.
\end{equation}
What we mean by the phrase ``$\Lambda$CDM-mimicking $f(R)$ gravity'' is some $f(R)$ model that reproduces a cosmic evolution that is the same as that of the General Relativistic $\Lambda$CDM model, namely
\begin{equation}\label{E_LCDM}
    E_{\Lambda\rm CDM}(z) = \frac{H_{\Lambda\rm CDM}(z)}{H_0} = \sqrt{\frac{2}{3}(1+q_0)(1+z)^3 + \frac{1}{3}(1-2q_0)^3}\,.
\end{equation}
Such an evolution is kinematically specified by the cosmographic condition $j=1$ \citep{Chakraborty:2022evc}. 
Let us define the background-related variables
\begin{equation}
E = \frac{H}{H_0}\,,\quad    x = \frac{\dot{f'}}{f'H}\,,\quad\Omega=\frac{\rho}{3f'H^2}\,,
\end{equation}
and the perturbation-related variables \citep{Carloni:2007yv,Ananda:2008tx}
\begin{equation}
   \Delta_m = \frac{a^2}{\rho_m} D^2 \rho_m\,,\quad \mathcal{R} = a^2 D^2 R\,,
\end{equation}
with $D^2=D_a D^a$ and $D_a$ being the projected covariant derivative on the spatial hypersurface orthogonal to the cosmological fundamental 4-velocity. In terms of the above variables, the perturbation equations for $\Lambda$CDM-mimicking $f(R)$ gravity, without using any quasistatic approximation, can be expressed as \citep{MacDevette:2024wpg}
\begin{subequations}\label{eq:full_cov}
\begin{align}
& E^2 \frac{d^2\Delta_m^{(k)}}{dz^2} + E^2\frac{(q+x)}{(1+z)} 
\frac{d\Delta_m^{(k)}}{dz} + 3E^2\frac{\Omega}{(1+z)^2} \Delta_m^{(k)} = \nonumber\\
&-\frac{ x }{2(q+1)(1+z)} \frac{d\hat{\mathcal{R}}^{(k)}}{dz} +\left( \frac{\frac{\hat{k}^2}{E^2}(1+z)^2 x+3 q (x-1)-3}{6(q+1)(1+z)^2}\right)\hat{\mathcal{R}}^{(k)},
\end{align}
\begin{align}
&\frac{d^2\hat{\mathcal{R}}^{(k)}}{dz^2} -\frac{ (q (x+4)+11 x-6 \Omega +4) }{x(1+z)} \frac{d\hat{\mathcal{R}}^{(k)}}{dz} +\nonumber\\
&\left(\frac{\hat{k}^2}{E^2}+\frac{2 q (q (x+2)+6 x-3 \Omega +10)-27 \Omega +16+34x}{x(1+z)^2}\right)\hat{\mathcal{R}}^{(k)}\nonumber\\
&=\frac{6 E^2 (q+1)}{(1+z)} \frac{d\Delta_m^{(k)}}{dz} -\frac{6E^2 (q+1) \Omega }{x(1+z)^2} \Delta_m^{(k)}, 
\end{align}
\end{subequations}
For a fully quasi-static approximation, this becomes
\begin{equation}\label{eq:full_QS}
    (1+z)^2 \frac{d^2{\Delta_m^{(k)}}}{dz^2} +(1+z)q \frac{d\Delta_m^{(k)}}{dz} =\Omega \left(\frac{2 \frac{\hat{k}^2}{h^2} x- 3 \frac{(q+1)}{(1+z)^2} }{\frac{\hat{k}^2}{h^2} x -2  \frac{(q+1)}{(1+z)^2}}\right)\Delta_m^{(k)}.
\end{equation}
It is to be noted that for GR, one has 
\begin{align}\label{GR_reduction}
\mathcal{R}^{(k)} =  \rho_m\Delta_m^{(k)} \,,
\end{align}
and hence the perturbative level is described by a single second-order differential equation for $\Delta_m$ \citep{MacDevette:2024wpg}. This is consistent with the fact that GR is a second-order theory.
Lastly, the growth function $S$ and the growth index parameter of matter perturbation $\gamma$, are defined as \citep{MacDevette:2024wpg}
\begin{eqnarray}
    S &\equiv& -(1+z) \frac{d \ln \Delta_m^{(k)}}{dz} = \Omega_m^{\gamma}\,,\label{eq:growthfunction}
    \\
    \gamma &=& \frac{\ln\left[-(1+z) \frac{d\ln \Delta_m^{(k)}}{dz}\right]}{\ln \Omega_m}\,.\label{eq:growthindexparam}
\end{eqnarray}
The growth index parameter $\gamma$, in particular, is a good discriminator between GR and $f(R)$ gravity at the level of matter perturbation. For the General Relativistic $\Lambda$CDM model, $\gamma\approx \frac{6}{11}$, whereas for $f(R)$ models, even the $\Lambda$CDM-mimicking ones, $\gamma$ is not only a variable in time, but also shows a dispersion (scale-dependence). 

While solving the perturbation equations \eqref{eq:full_cov} and \eqref{eq:full_QS} , the initial conditions must be set carefully. In this analysis, we adopt a present-day deceleration parameter of $q_0 = q(z=0) \approx -0.55$ and establish our initial conditions at a high redshift of $z_{\rm in} = 1000$. Because these initial conditions are fundamentally integrated into our subsequent 21\,cm survey analysis, their precise specification is essential. At $z=1000$, $f(R)$ gravity is virtually indistinguishable from GR, meaning the theory effectively mimics a standard $\Lambda$CDM cosmology. Relying on this equivalence, we use the \textsc{python} package \texttt{astropy} to calculate the initial dimensionless Hubble parameter $E_{\rm in}$ and the initial deceleration parameter $q_{\rm in}$. The corresponding initial density parameter, $\Omega_{\rm in} \equiv \Omega_{\Lambda{\rm CDM}}$, is similarly computed via \texttt{astropy} in conjunction with the relation provided by \cite{MacDevette:2024wpg}. The quantity $x_{\rm in}$ parametrises the initial deviation from GR. It is strategically chosen to be extremely small, ensuring the model accurately reproduces the behaviour of GR at early epochs, yet not so small that it becomes mathematically equivalent to it. The choice of $x_{\rm in}$ is not fixed \emph{a priori}; rather, it is determined via a trial-and-error approach. Our objective is to isolate a value that is fully consistent with the observational parameters of 21\,cm survey analyses, specifically for \textsc{SKA} and \textsc{PUMA}. For these surveys, we adopt a common and conservative range of  modes, $k \sim 10^{-2}$--$10^{-1}\,{\rm Mpc}^{-1}$. 

It is a well-established feature of $f(R)$ gravity that the growth index $\gamma(k,z)$ exhibits both a time and scale dependence, varying with redshift $z$ and wave number $k$, as explicitly demonstrated in equation~(\ref{eq:growthindexparam}). By evaluating $\gamma(k,z)$ across the aforementioned range of scales, we are able to constrain $x_{\rm in}$ to appropriately chosen values that guarantee a physically viable (i.e., strictly positive) growth index. Following the computation of all these initial parameters based on the $\Lambda$CDM limit, the resulting values are summarised below:
\begin{equation}
\quad\quad
\begin{aligned}
x_{\rm in} &= -10^{-14} \,,\quad \Omega_{m,\rm in}= 0.999999997 \,,\\
E_{\rm in} &= 17346.495 \,,\quad q_{\rm in} = 0.499999996 \,.
\end{aligned}
\label{eq:initial condition parameters}
\end{equation}

\begin{figure}
    \centering
    \includegraphics[width=\linewidth]{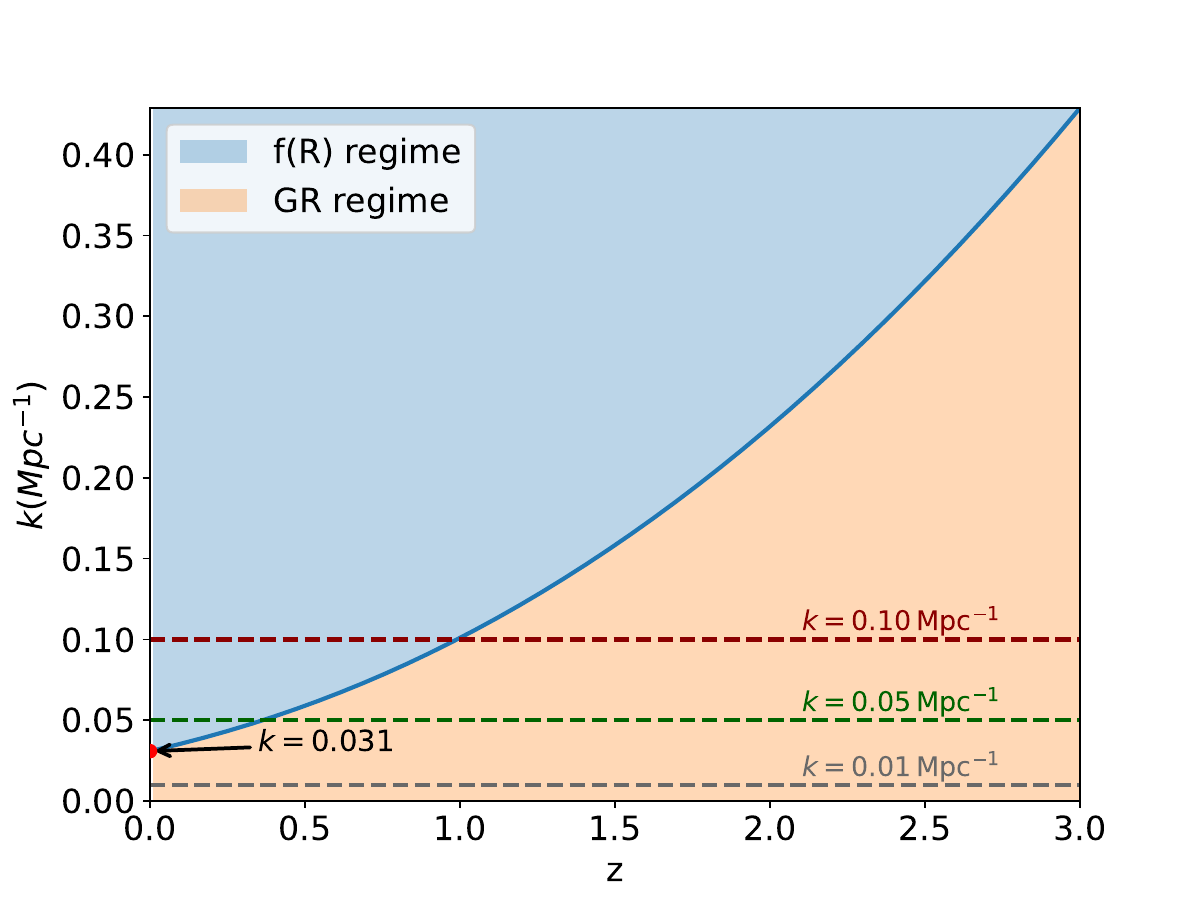}
    \caption{Evolution of the transition redshift (solid blue line) under the quasi-static approximation, given the initial conditions in equation~(\ref{eq:initial condition parameters}). This curve delineates the GR (blue) and $f(R)$ gravity (light orange) regimes. At $z=0$, the transition occurs at a Fourier wave number $k_0 \simeq 0.031\,\mathrm{Mpc}^{-1}$. Dotted lines highlight specific scales at $k = 0.01\,\mathrm{Mpc}^{-1}$ (grey), $0.05\,\mathrm{Mpc}^{-1}$ (green), and $0.1\,\mathrm{Mpc}^{-1}$ (red), which are the focus of the remaining analysis.}
    \label{fig:Fr_GR_region}
\end{figure}
For the perturbations, the initial conditions are set based on constraints from the cosmic microwave background (CMB), from which we obtain the initial density contrast $\Delta_{\rm in}$. We then solve equation~(\ref{eq:growthindexparam}) by adopting $\gamma \simeq 6/11$ at $z=1000$. This allows us to consistently fix the initial values of $\Delta_{\rm in}$ and its temporal derivative $\Delta'_{\rm in}$, yielding
\begin{equation}
\quad\quad
\Delta_{\rm in} = 1\times10^{-5} \,, \quad
    \Delta'_{\rm in} = \left(\frac{d\Delta_m}{dz}\right)_{\rm in} =  -9.99\times 10^{-9} \,.
\label{eq:delta initial}
\end{equation}
This ensures that $\Delta_{\rm in}$ sets a small initial matter over-density consistent with linear theory and $\Delta'_{\rm in}$ enforces the expected growing mode scaling at high redshift \citep{Ma1995}. 

Following the determination of the background quantities, we compute the initial conditions for the spatial variations of the Ricci scalar, $\hat{\mathcal{R}}^{(k)}_{\rm in}$, and its temporal derivative $\hat{\mathcal{R}}^{(k)\prime}_{\rm in}=(d\hat{\mathcal{R}}^{(k)}/dz)_{\rm in}$, by demanding that at such a high redshift the underlying $f(R)$ theory is virtually indistinguishable from GR. Hence, Eq.\eqref{GR_reduction} gives us the following expressions:
\begin{equation}\label{eq: temporal fluctuations}
    \hat{\mathcal{R}}^{(k)} = 3(1-3w)E^2\,\Omega_m\Delta_m^{(k)} \,,
\end{equation}
and
\begin{equation}
    \hat{\mathcal{R}}^{(k)\prime} = 3(1-3w)E^2\,\Omega_m\Delta_m^{(k)}\left[3+(1+z)\frac{\Delta_m^{(k)\prime}}{\Delta_m^{(k)}}\right] \,.
\end{equation}
To evaluate these expressions at the initial redshift $z_{\rm in} = 1000$, we apply the standard assumption of a matter-dominated epoch, setting the equation-of-state parameter to $w=0$. At this early time, the model resides deep within the GR regime ($f' \approx 1$); consequently, the matter density parameter converges to the total density parameter, such that $\Omega_m(z=1000) \approx \Omega(z=1000)$. Furthermore, the dimensionless Hubble parameter reduces to the standard $\Lambda$CDM limit, $E = E_{\Lambda\rm CDM}$, which is evaluated at $z=1000$ as a specified value in (\ref{eq:initial condition parameters}). By substituting these background values alongside the previously derived initial matter density perturbations, $\Delta_m^{(k)}$, and their derivatives, $\Delta_m^{(k)\prime}$ (\ref{eq:delta initial}), we obtain the complete set of initial conditions for $\hat{\mathcal{R}}^{(k)}$ and $\hat{\mathcal{R}}^{(k)\prime}$. The initial curvature perturbation tracks the density perturbation and remains slowly varying at early times, before modified-gravity effects become significant. 

\begin{figure*}
\centering
\begin{subfigure}{0.42\textwidth}
    \centering
    \includegraphics[width=\linewidth]{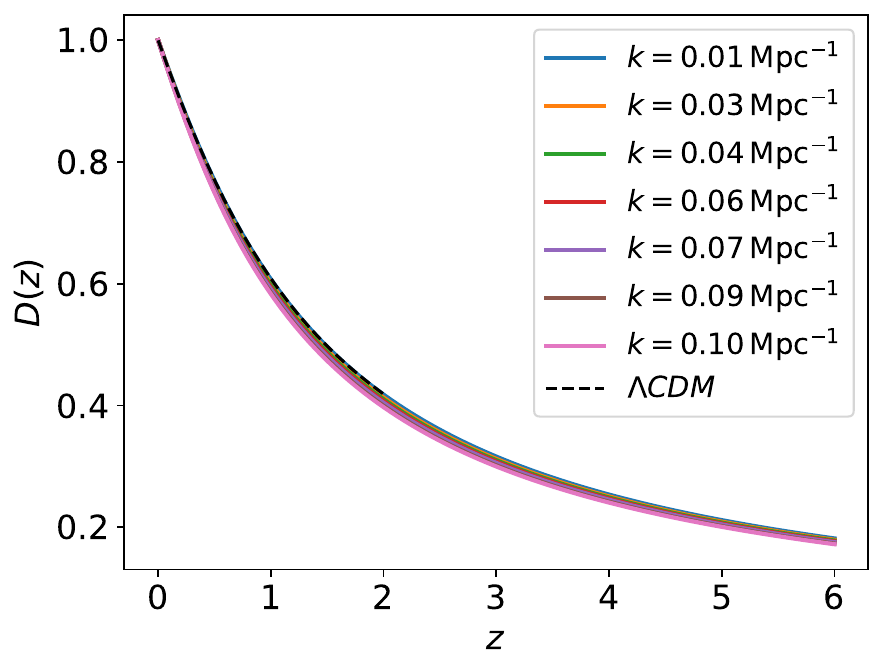}
    \caption{}
\end{subfigure}
\begin{subfigure}{0.42\textwidth}
    \centering
    \includegraphics[width=\linewidth]{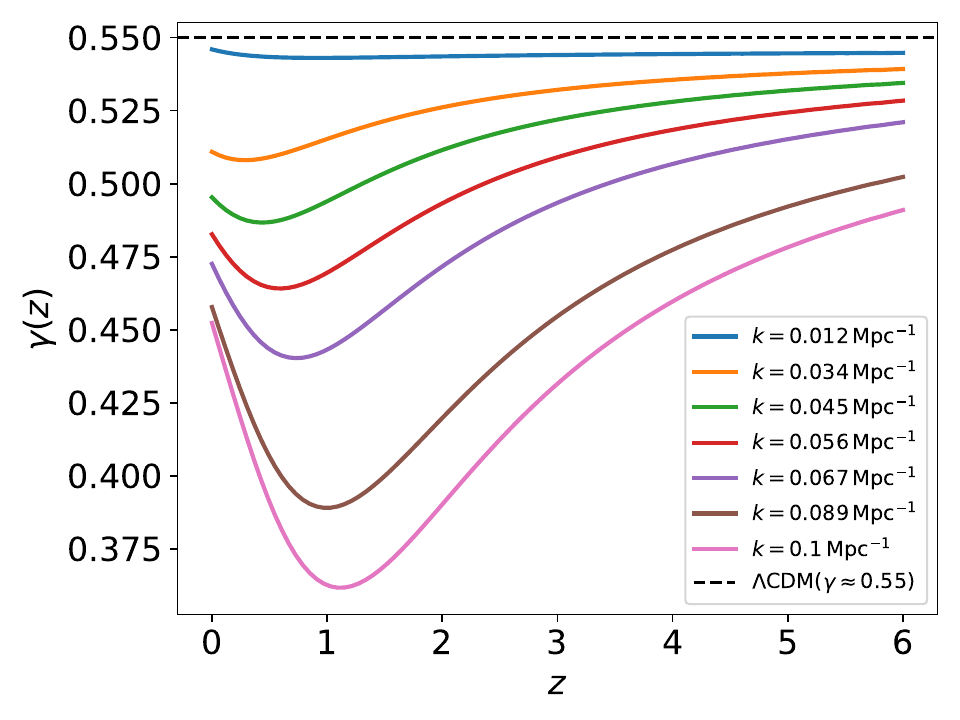}
    \caption{}
\end{subfigure}
    \caption{In these plots, we show the evolution of different modes and the variation of the growth index within the full quasi-static approximation (\ref{eq:full_QS}) using the initial conditions discussed above. The plots also highlight the relevant modes probed by 21\,cm surveys, which are important for understanding the modes responsible for galaxy formation.}
    \label{fig:full growth}
\end{figure*}

\begin{figure*}
\centering
\begin{subfigure}{0.42\textwidth}
    \centering
    \includegraphics[width=\linewidth]{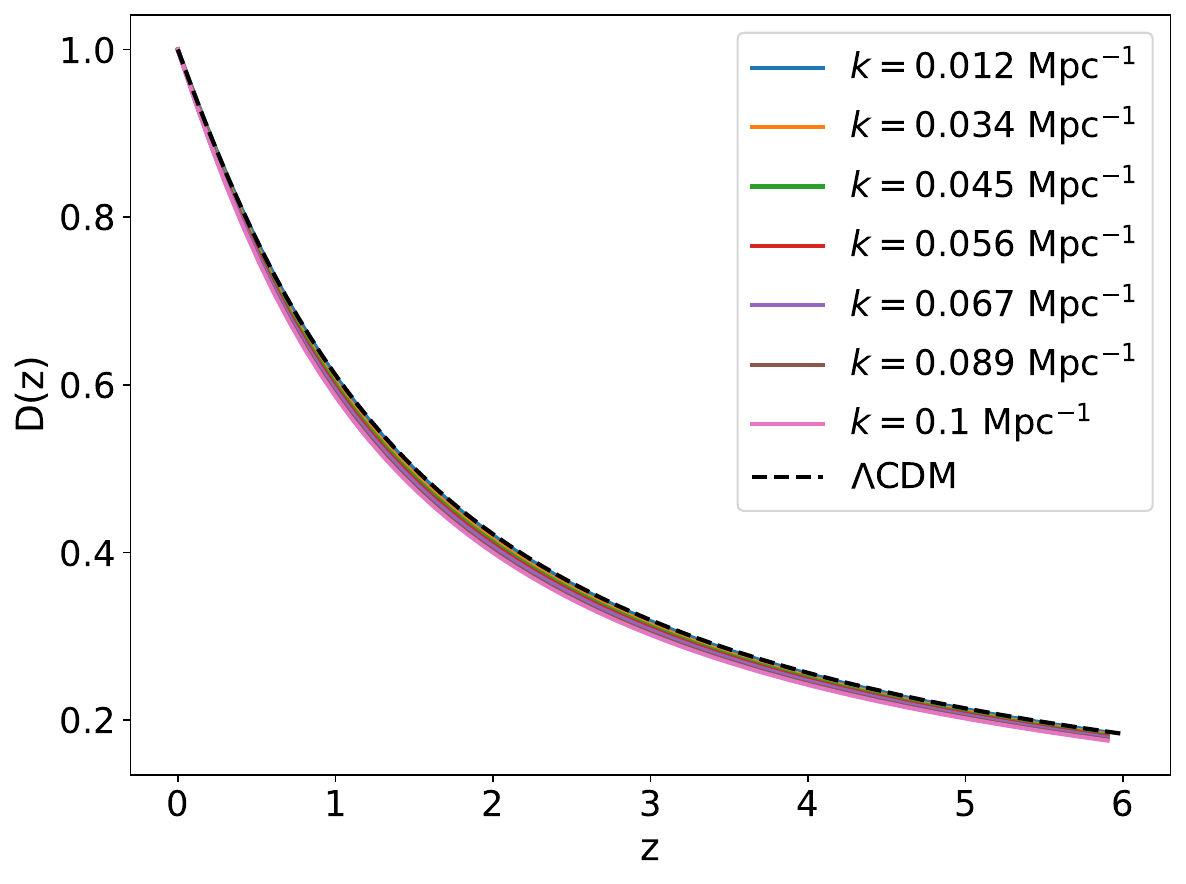}
    \caption{}
\end{subfigure}
\begin{subfigure}{0.42\textwidth}
    \centering
    \includegraphics[width=\linewidth]{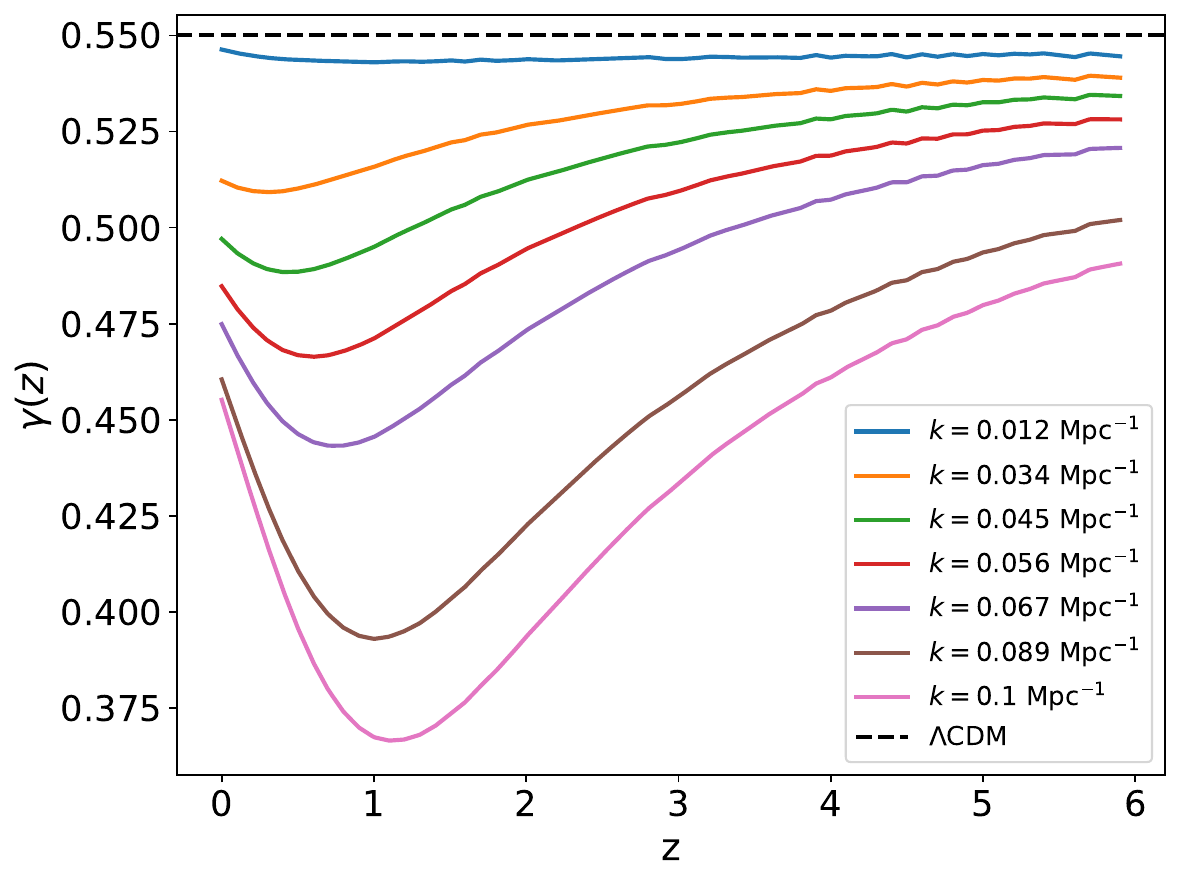}
    \caption{}
\end{subfigure}
    \caption{In these plots, we show the evolution of different modes and the variation of the growth index within the exact equation (\ref{eq:full_cov}) using the initial conditions discussed above. The plots also highlight the relevant modes probed by 21\,cm surveys, which are important for understanding the modes responsible for galaxy formation.}
    \label{fig:exact growth}
\end{figure*}

We evolve both the growth factor and the growth index using the full quasi-static (QS) approximation and the exact method. Throughout this paper, the normalized matter density perturbation is defined as 
\begin{equation}
\quad\quad
D(z)=\frac{\Delta_m(k,z)}{\Delta_m(k,0)}\,.    
\end{equation}
As shown in Figs. (\ref{fig:full growth}) and (\ref{fig:exact growth}), a realistic $21$\,cm survey requires an appropriate range of redshift $z$, wave number $k$, and a positive $\gamma(k,z)$. The resulting plots are consistent when we choose $x_{\rm in}=-10^{-14}$. Fig.~(\ref{fig:Fr_GR_region}) illustrates the epochs at which various Fourier modes transition into the $f(R)$ gravity regime. As previously discussed, our 21-cm survey analysis targets the scale range $k \sim 10^{-2}$--$10^{-1}\,\mathrm{Mpc}^{-1}$. At the present epoch ($z=0$), the transition scale corresponds to $k_0 \simeq 0.031\,\mathrm{Mpc}^{-1}$, meaning all modes with $k > k_0$ have already entered the $f(R)$ regime. We specifically highlight two such modes at $k = 0.05\,\mathrm{Mpc}^{-1}$ (green dotted line) and $0.1\,\mathrm{Mpc}^{-1}$ (red dotted line), whereas the larger-scale $k = 0.01\,\mathrm{Mpc}^{-1}$ mode remains within the GR regime. 

\begin{figure*}
\centering
\begin{subfigure}{0.42\textwidth}
    \centering
    \includegraphics[width=\linewidth]{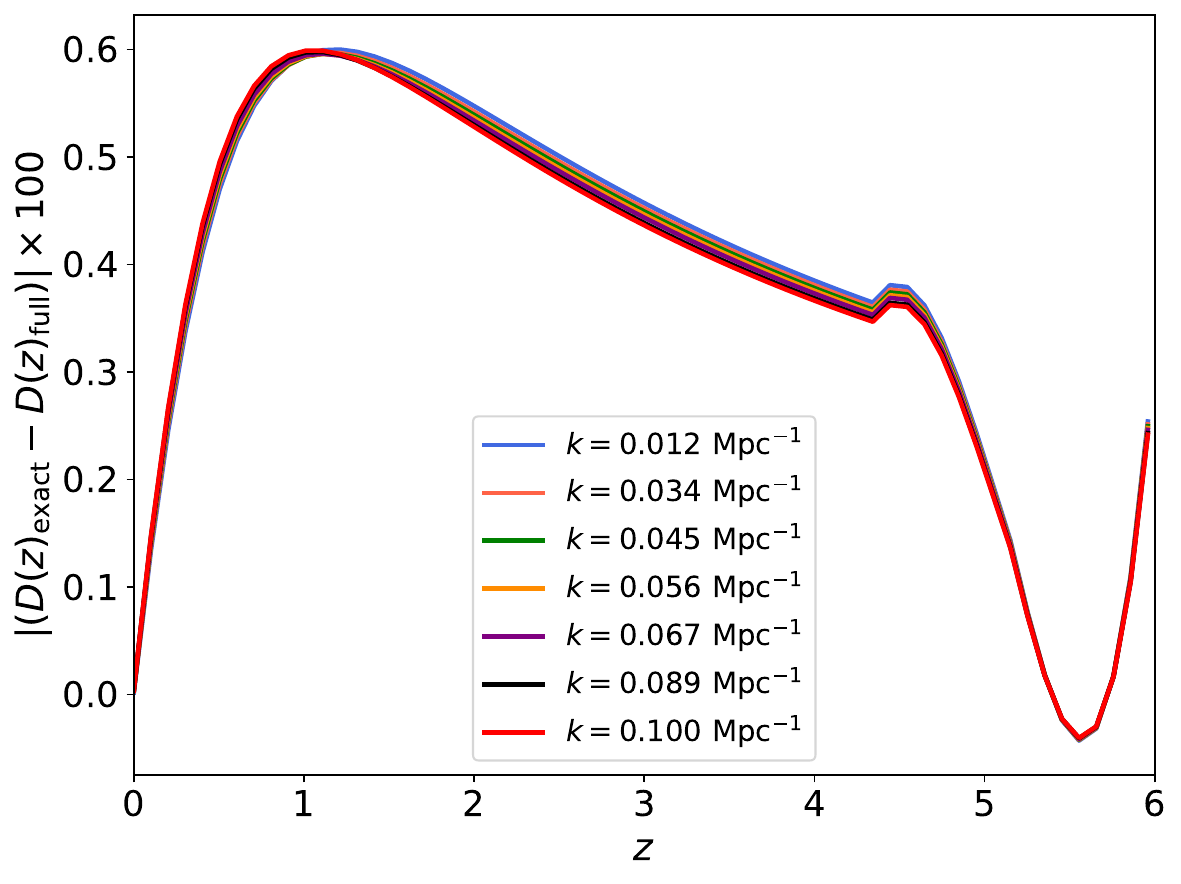}
    \caption{}
\end{subfigure}
\begin{subfigure}{0.42\textwidth}
    \centering
    \includegraphics[width=\linewidth]{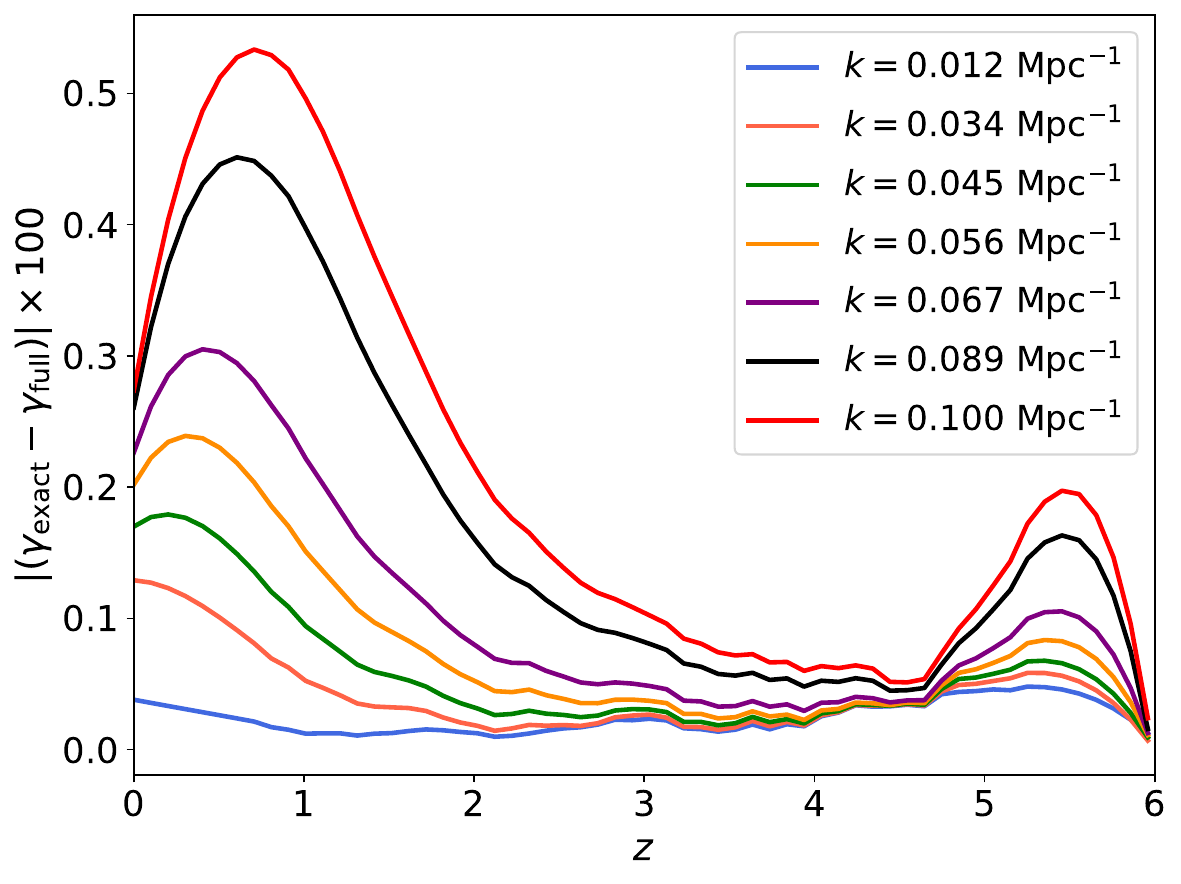}
    \caption{}
\end{subfigure}
    \caption{ Absolute percentage difference between the exact solutions and the full QS (solid) solutions for: (a) the normalised perturbation D(z), and (b) the growth index, $\gamma(k,z)$.}
    \label{fig:parsentage different}
\end{figure*}

\begin{figure*}
\centering
\begin{subfigure}{0.45\textwidth}
    \centering
    \includegraphics[width=\linewidth]{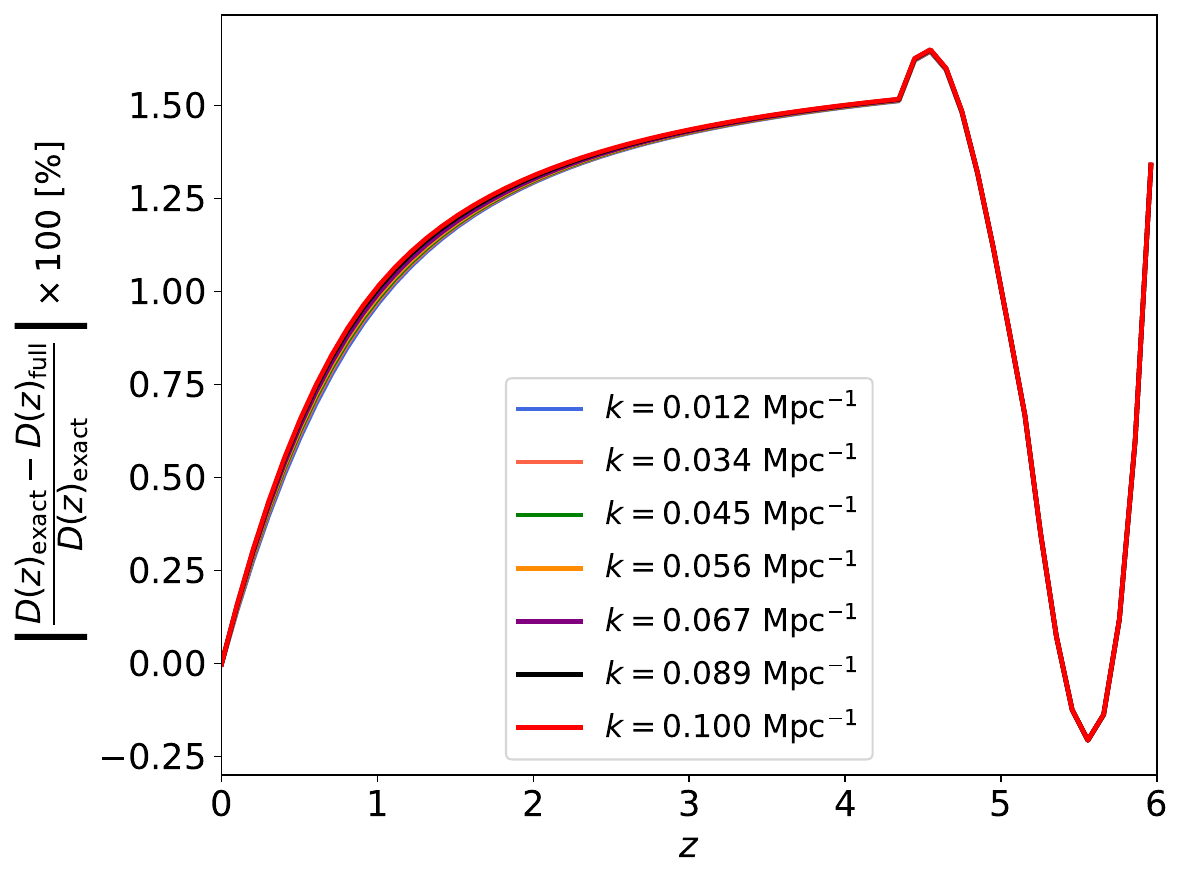}
    \caption{}
\end{subfigure}
\begin{subfigure}{0.45\textwidth}
    \centering
    \includegraphics[width=\linewidth]{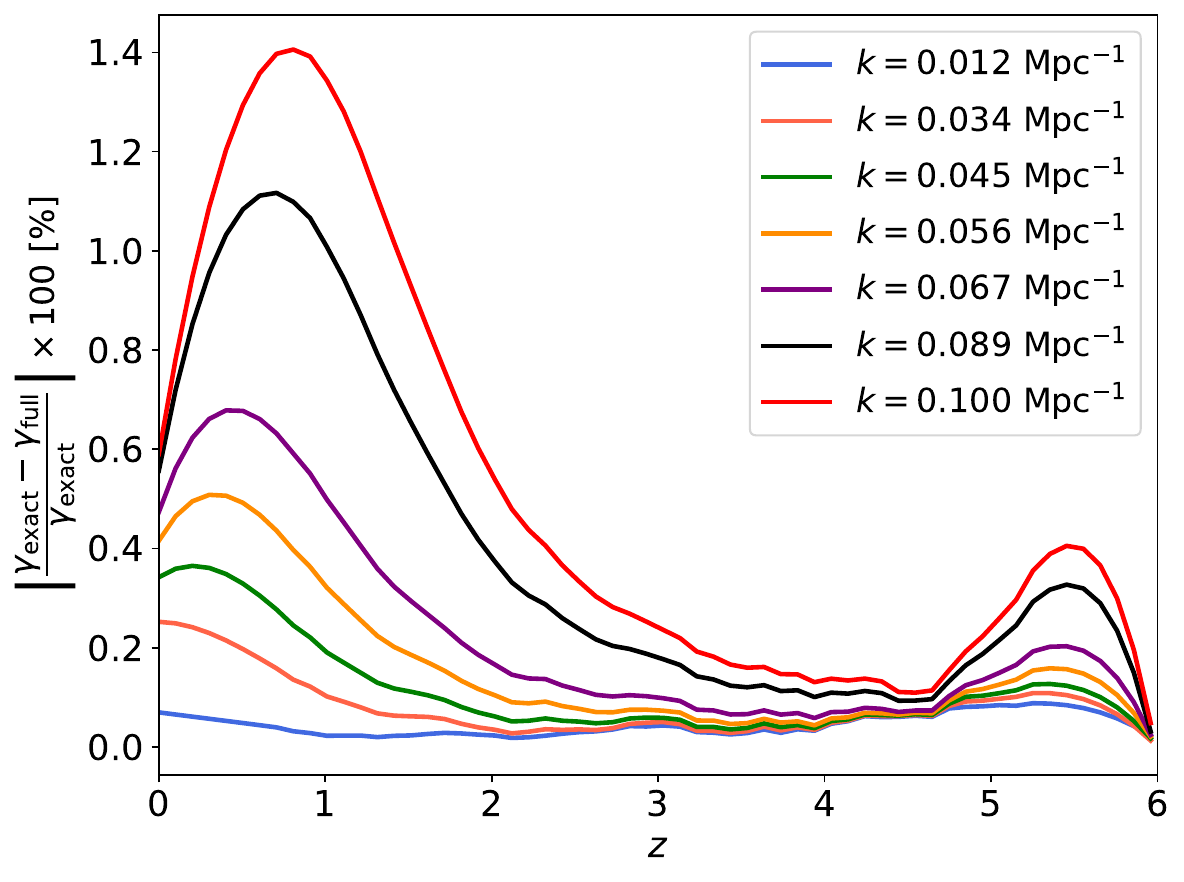}
    \caption{}
\end{subfigure}
    \caption{Relative percentage difference between the exact solutions and the full QS (solid) solutions for: (a) the normalised perturbation D(z), and (b) the growth index, $\gamma(k,z)$.}
    \label{fig: relative parsentage different}
\end{figure*}

Fig. (\ref{fig:parsentage different}) and Fig.(\ref{fig: relative parsentage different}) illustrates the Absolute and Relative percentage difference between the exact and full QS methods as a function of redshift, $z$, across various wave numbers, $k$. The deviation is particularly pronounced in the evolution of the growth index, $\gamma(k,z)$. For scales $k > 0.031\,\mathrm{Mpc}^{-1}$, which have entered the $f(R)$ regime, the percentage difference exhibits a slight increase, reflecting the distinct dynamical treatments of the two approaches. Conversely, at lower wavenumbers that remain within the GR-dominated regime, the two methods yield nearly identical results. The negligible difference at these larger scales confirms that both theoretical formulations consistently recover the standard GR limit.

In the later sections, we will demonstrate how 21-cm observations across these specific scales can be utilised to isolate $f(R)$ signatures and quantify deviations from standard GR. In this analysis, we restrict ourselves to $k \leq 0.1\,{\rm Mpc}^{-1}$ while larger values, which are not relevant for the present study, may lead to negative values of $\gamma(k,z)$. Having established the consistency of the model, we next aim to constrain the growth index $\gamma(k,z)$ and the bias parameter $\beta(k,z)$ using observations from future $21$\,cm radio surveys.


\section{21cm Cosmology}\label{sec:21cm_Basics}
In the era of precision cosmology, cosmological models are tightly constrained using observations from the CMB and large-scale structure. In addition to next-generation galaxy number count surveys, intensity mapping experiments will target the integrated 21-cm spectral line emission from neutral hydrogen, the most abundant element in the Universe. These surveys do not resolve individual galaxies and can therefore rapidly cover large cosmic volumes while providing extremely accurate red shifts directly from the frequency dimension. The 21-cm signal serves as a powerful cosmological probe. It enables exploration of the cosmic dark ages and cosmic dawn \citep{Ali-Haimoud:2013hpa,Mondal:2023xjx,Saurabh_RRI:2021mxo}, studies of dark matter physics \citep{Darkmatter_21cm:2023slb,Darkmatter_21cm_2:2016sur}, and tests of GR on cosmological scales \citep{Hall:2012wd} and the effects of early universe \citep{Joshi:2025swr}.
\subsection{A brief review of the Physics of 21cm Line}
The 21 cm line or HI line corresponds to the hyperfine transition between two closely spaced energy levels of the hydrogen atom. These energy levels are known as \textit{singlet} and \textit{triplet} states.  
The relative abundance of the singlet ($n_0$) and triplet states ($n_1$) can be written as \citep{21cm_BOOK, 21cm_review},
\begin{equation}
\quad\quad
    \frac{n_1}{n_0}=\frac{g_1}{g_0}\exp{\left(-\frac{T_*}{T_S}\right)}\;, \label{eq:level-occu-ratio}
\end{equation}
where $g_1/g_0=3$ is the degeneracy of levels, $T_*=hc/{k_\mathrm{B}\lambda_{21}} = 0.068$K, is the temperature corresponding to the 21cm emission in the atom's rest frame. The spin temperature $T_S$ decides the abundance of singlet and triplet states and couples with different temperatures depending on the underlying physics \citep{Pritchard_2012}. 
At high redshifts ($z<200$) it couples with kinetic temperature $T_K$ of the gas. At redshift ($z<40$), due to the expansion of the universe, gas temperature density decreases, collision becomes ineffective and $T_S$ couples with the CMB temperature $T_{\gamma}$. After the formation of first luminous objects, $T_S$ couples with the Lyman $\alpha$ photons which correspond to the colour temperature $T_{\alpha}$  \citep{21cm_review,Pritchard_2012, Bharadwaj:2004nr,Ankita_Bera:2022vhw}.
The observable quantity is the excess brightness temperature relative to the CMB and can be expressed as
\begin{equation}
\quad\quad
\delta T_\mathrm{b}= \frac{T_S- T_\gamma}{1+z}\tau\;,
\end{equation}
where $\tau\ll 1$ is the reionisation depth. After plugging $\tau$ value, we finally arrive at \citep{Pritchard_2012}
\begin{equation}\label{eq:Temp_Fluc_21cm}
\begin{aligned}
\delta T_\mathrm{b} \approx 27\left(\frac{\Omega_{\mathrm{b},0}h^2}{0.023}\right)\left(\frac{0.15}{\Omega_{\mathrm{m},0}h^2}\frac{1+z}{10}\right)^{1/2}{x}_\mathrm{HI}(1+\delta_\mathrm{b})\left(\frac{T_S-T_{\gamma}}{T_S}\right)\\\left[\frac{\partial_rv_r}{(1+z)H(z)}\right]\ \mathrm{mK}\;.
\end{aligned}
\end{equation} 
In this equation, $\Omega_{\mathrm{b},0}$ and $\Omega_{\mathrm{m},0}$ are the present values of baryon and total matter density parameters. Neutral fraction of hydrogen is written as $x_\mathrm{HI}$, $\delta_\mathrm{b}$ stands for fractional overdensity in baryons. The last term in the square brackets is the correction due to peculiar velocities along the line of sight (LoS). 
\subsection{Surveys of HI Intensity Mapping \label{sec:HISurvey_IntenMapp}}
To probe the 21cm signal, we employ radio telescopes in two different modes of operation: 
\begin{enumerate}
\item Single Dish Mode (SD): In this mode of operation, we simply add auto-correlation signals from single dishes. After this, we take the Fourier transform. This mode allows us to probe relatively small $k$ values. SKA-Mid is optimised to operate in this mode.
\item Interferometer Mode (IF): In this case, cross-correlated signals from the array elements are combined. This allows for high resolution on small angular scales. We directly get the Fourier transform of the sky.  PUMA is designed to operate in this mode.
\end{enumerate}
For line intensity mapping surveys, the thermal noise from the instrument dominates the noise component on the scales of interest, whereas, the shot noise contribution remains minuscule \citep{Gong:2011qf}. So we neglect the short noise contribution for the purpose of this paper  \citep{Karagiannis:2024noise, Santos:2015noise, Bull:2014rha}. 
For surveys in the SD mode, the noise power spectrum of the instrument is \citep{Santos:2015gra}
\begin{equation}
    P_\mathrm{N}^\mathrm{SD}(k_\perp,z) = T^2_\mathrm{sys}(z)\, \chi^2(z)\, \lambda(z)\, \frac{1+z}{H(z)}\, \frac{4\pi f_\mathrm{sky}}{\eta\, N_\mathrm{pol}\, N_\mathrm{dish}\, t_\mathrm{survey}\, \beta^2_\perp(k_\perp, z)}\;. \label{eq:Noise_Power_SD}
\end{equation}
Here the system temperature $T_\mathrm{sys}$ is taken from \cite{SKA:2018ckk}, $\chi$ and $H(z)$ are respectively the co-moving distance and Hubble parameter. For SKA-Mid, which is the only SD mode survey considered in this paper, we take efficiency $\eta=1$ and the number of polarisations per feed $N_\mathrm{pol}=2$ following \cite{CosmicVisions21cm:2018rfq}. Further, $\lambda(z)= \lambda_{21}(1+z)$ is the redshifted wavelength of 21cm line and $\lambda_{21}=21$cm is the wavelength in the rest frame of the emitter atom. Also, $\beta_\perp(k_\perp,z)$ is the transverse effective beam which in the Fourier space is \citep{Bull:2014rha}
\begin{equation}
\quad
\beta_\perp(k_\perp,z) = \exp\left(-\frac{k_\perp^2 \chi^2(z) \theta_\mathrm{b}^2(z)}{16\ln 2}\right)\;,
\end{equation}
with $\theta_\mathrm{b}(z)=1.22\lambda(z)/D_\mathrm{dish}$ being the FWHM of individual dish beam.
It turns out that the effective beam in the radial direction may be neglected on account of very high frequency resolution of IM experiments \citep{Bull2015_21cm}. On the other hand, 
the noise power spectrum for interferometer mode is given by \citep{CosmicVisions21cm:2018rfq}
\begin{equation}
    P_\mathrm{N}^\mathrm{IF}(k_\perp,z) = T_\mathrm{sys}^2(z)\,\chi^2(z)\,\lambda(z)\,\frac{(1+z)}{H(z)}\left(\frac{\lambda^2(z)}{A_\mathrm{e}}\right)^2 \frac{2\pi f_\mathrm{sky}}{t_\mathrm{survey}\, n_\mathrm{b}(u,z)\,\theta_\mathrm{b}^2} \label{eq:Noise_Power_IF}\;.
\end{equation}
Here $T_\mathrm{sys}$ for PUMA is  taken from \cite{CosmicVisions21cm:2018rfq} and for HIRAX it is the sum of receiver temperature $T_\mathrm{rx}=50$K and sky temperature is taken to be the same as PUMA, the effective area $A_\mathrm{e} = \eta\pi(D_\mathrm{dish}/2)^2$ depends on the efficiency $\eta$. For both HIRAX and PUMA, we take $\eta=0.7$ and $n_\mathrm{b}(u)$ is the baseline number density in the uv plane. Its expression for both HIRAX and PUMA is given in 
Appendix D of \cite{CosmicVisions21cm:2018rfq}. The total survey area is calculated for PUMA using Table \ref{table:Radio_arry}.
\begin{table}
    \centering
    \begin{tabular}{l c c c}
    \hline\hline
     & PUMA & SKA-Mid \\
        \hline
        Redshift range & $0.3 - 6.0$ & $0.35 - 3.05$\\
        Integration time, $t_\mathrm{survey}$ (sec) & $1.58\times 10^8$ & $3.6\times 10^7$ \\
        Sky fraction, $0.36$ & $0.5$  & $0.49$ \\
        Dish diameter, $D_\mathrm{dish}$ (m) & 6 & 15 \\
        Maximum baseline & 1.0 km & 150 km \\
        $N_\mathrm{dish}$ & 32,000 & 197 \\
        \hline\hline
    \end{tabular}
    \caption{The instrumental details of PUMA radio interferometers used in this work \citep{CosmicVisions21cm:2018rfq}. Information about SKA-Mid (Band 1) is based on data from the official website \href{https://www.skao.int/en/explore/telescopes/ska-mid}{SKA-MID}.}
    \label{table:Radio_arry}
\end{table}
Cosmological constraints are usually given with the help of power spectrum which is the Fourier transform of the two-point correlation function (2PCF). When the primordial curvature perturbation is considered as Gaussian random field and its evolution is assumed to be linear, the 2PCF is sufficient to extract all the relevant information. In such a case, it is known that all odd correlations are zero and all even correlators are expressible in terms of the 2PCF. 

\subsection{Power Spectrum of 21cm Intensity}
The 21cm power spectrum $P_{21}(\mathbf{k},z)$ is defined through 
2PCF of Fourier coefficient $\delta T_\mathrm{b}(\mathbf{k}, 
z)$ of the excess brightness temperature $\delta T_\mathrm{b}(\hat{\mathbf{n}},z)$, (c.f. \eqref{eq:Temp_Fluc_21cm})
\begin{equation}
\langle \delta T_\mathrm{b}(\mathbf{k},z)\delta T_\mathrm{b} (\mathbf{k^{\prime}},z)\rangle = (2\pi)^3P_{21}(\mathbf{k},z)\delta_{D}(\mathbf{k}+\mathbf{k^{\prime}}) \label{eq:21cm_Pow_Spec}
\end{equation}
assuming the LoS is along $\hat{z}$ axis and primordial anisotropies are absent, the expression for 21cm differential brightness temperature power spectrum $P_{21}$ is 
\begin{equation}
P_{21}(\mathbf{k},z)=\Bar{T}^2_\mathrm{b}(z)D_\mathrm{FoG}(\mathbf{k},z)(1+\mu^{2}\beta(k,z))^{2}b_{1}^{2}(z)P_m(k,z) + P_\mathrm{N}(\mathbf{k},z) \;,\label{eq:power_sp_eq}
\end{equation}
where $\bar{T}_\mathrm{b}$ is the mean brightness temperature, the term $D_\mathrm{FoG}(\mathbf{k}) = \text{exp}[-k^2\mu^2\sigma^2_\mathrm{P}(z)]$ corresponds to the \textit{Fingers of God} (FoG) effect, with $\sigma_\mathrm{P}$ denoting the LoS dispersion in relative galaxy velocities. In this paper, we use $\sigma_\mathrm{p}=350/H(z)$, as per 
\cite{FoG:1996cd}. Further,  $b$ is the neutral hydrogen bias given by \cite{Jolicoeur:2020eup}
\begin{equation}
    b(z) = 0.754 + 0.0877z + 0.0607z^2-0.00274z^3 \label{eq:b1_bias}
\end{equation}
and $\mu= \hat{k}\cdot \hat{z}$ is the cosine of angle {between $\hat{k}$ and} LoS, $\beta(k,z) = f(k,z/b(z))$, where $f$ is the linear growth rate related to growth factor $D$ by $f=d\ln{D(a)}/d\ln{a}$. The quantity, $P_\mathrm{m}({k},z)$ denotes the linear matter power spectrum. As was discussed before, the noise power spectrum $P_\mathrm{N}(\mathbf{k},z)$ doesn't get any contribution from the shot noise.   The only contribution comes from the instrumental noise $P_\mathrm{N}(\mathbf{k},z)$ with the expressions given in \eqref{eq:Noise_Power_SD} and \eqref{eq:Noise_Power_IF} for the single dish and interferometer modes of operation. 
The matter power spectrum $P_\mathrm{m}$ appearing in \eqref{eq:power_sp_eq} is related to the power spectrum of the primordial curvature perturbations $P_\zeta$ as \citep{CosmicVisions21cm:2018rfq} 
\begin{equation}
    P_\mathrm{m}(k,z) = \mathcal{T}^2(k,z)P_{\zeta}(k)\;,
\end{equation}
where $\mathcal{T}(k)$ is the matter transfer function. After introducing the $21$\,cm formalism, Eq.~\ref{eq:power_sp_eq} gives the corresponding $21$\,cm power spectrum. This expression depends on two important quantities, namely the growth and bias parameters, both of which are determined by the underlying gravity theory. In this work, we adopt the framework of $f(R)$ gravity and examine how it modifies the cosmological evolution. In the next section, we use the Fisher matrix formalism to constrain these two parameters, which is the main objective of this analysis. We then compare our results with the standard GR case using future $21$\,cm survey observations.
\section{Results and Discussion}\label{sec:results_and_discussion}
\begin{figure*}
\centering

\begin{subfigure}{0.33\textwidth}
    \centering
    \includegraphics[width=\linewidth]{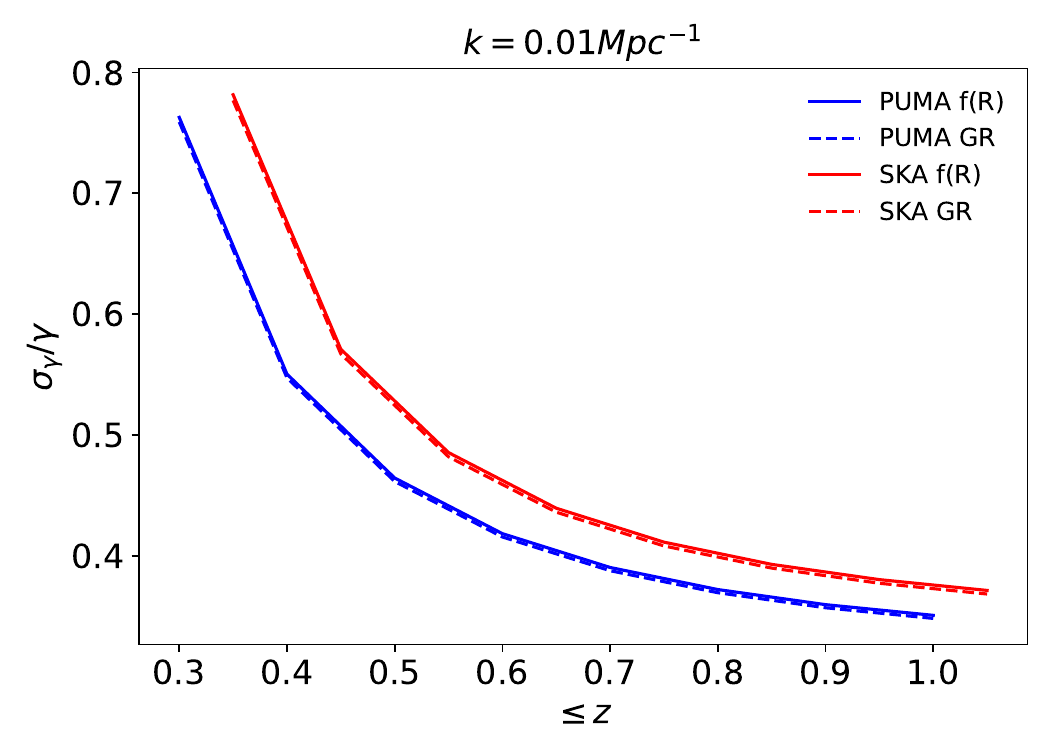}
    \caption{}
\end{subfigure}
\begin{subfigure}{0.33\textwidth}
    \centering
    \includegraphics[width=\linewidth]{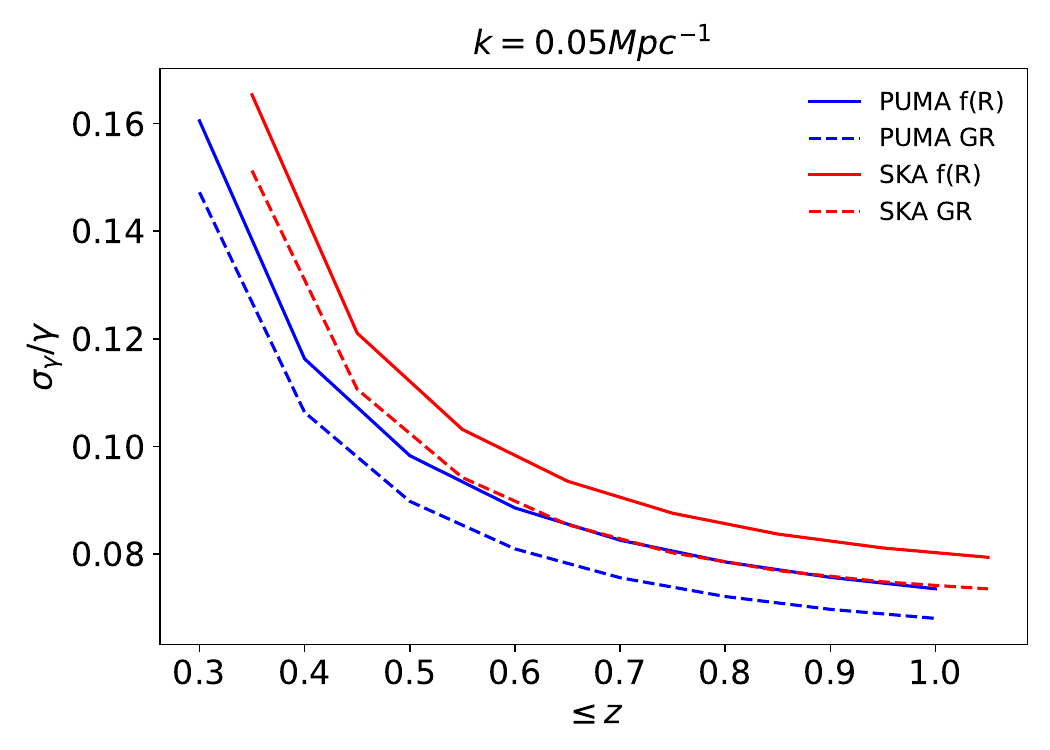}
    \caption{}
\end{subfigure}
\begin{subfigure}{0.33\textwidth}
    \centering
    \includegraphics[width=\linewidth]{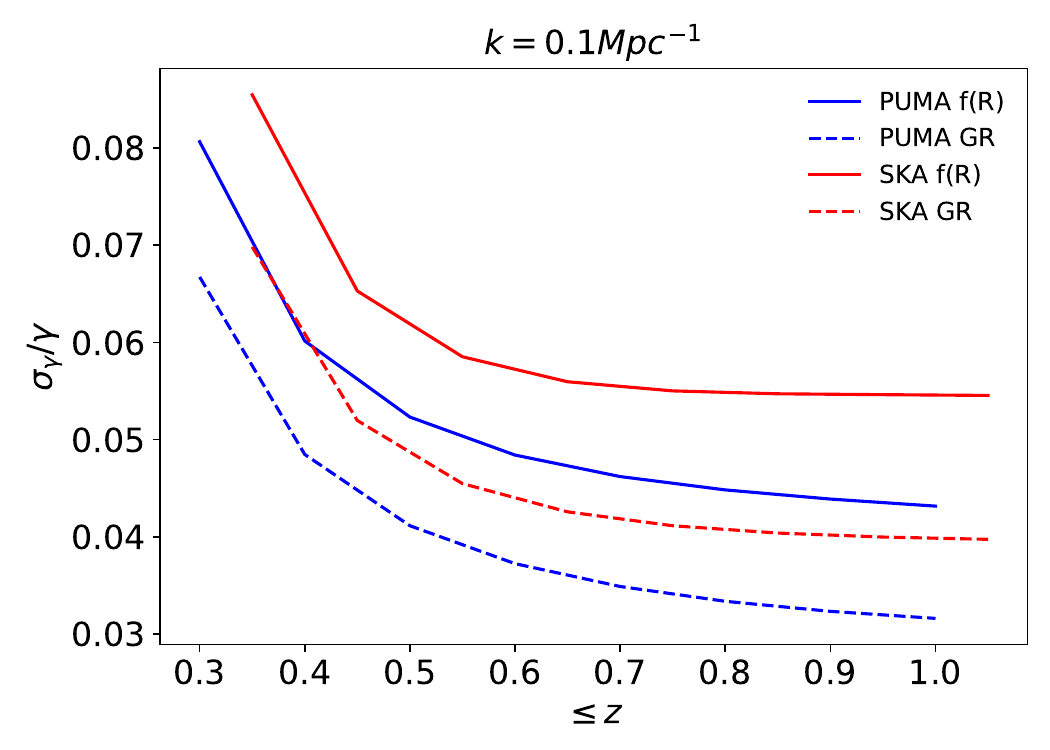}
    \caption{}
    \label{fig:gamma_full_error}
\end{subfigure}

\vspace{0.2cm}

\begin{subfigure}{0.33\textwidth}
    \centering
    \includegraphics[width=\linewidth]{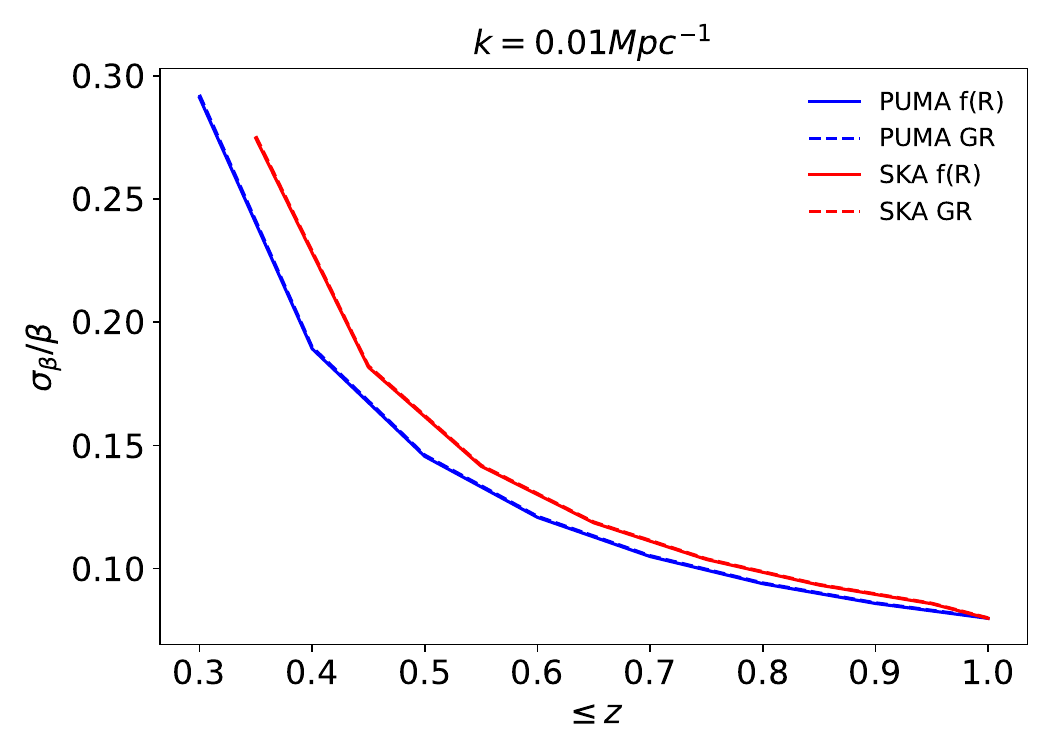}
    \caption{}
\end{subfigure}
\begin{subfigure}{0.33\textwidth}
    \centering
    \includegraphics[width=\linewidth]{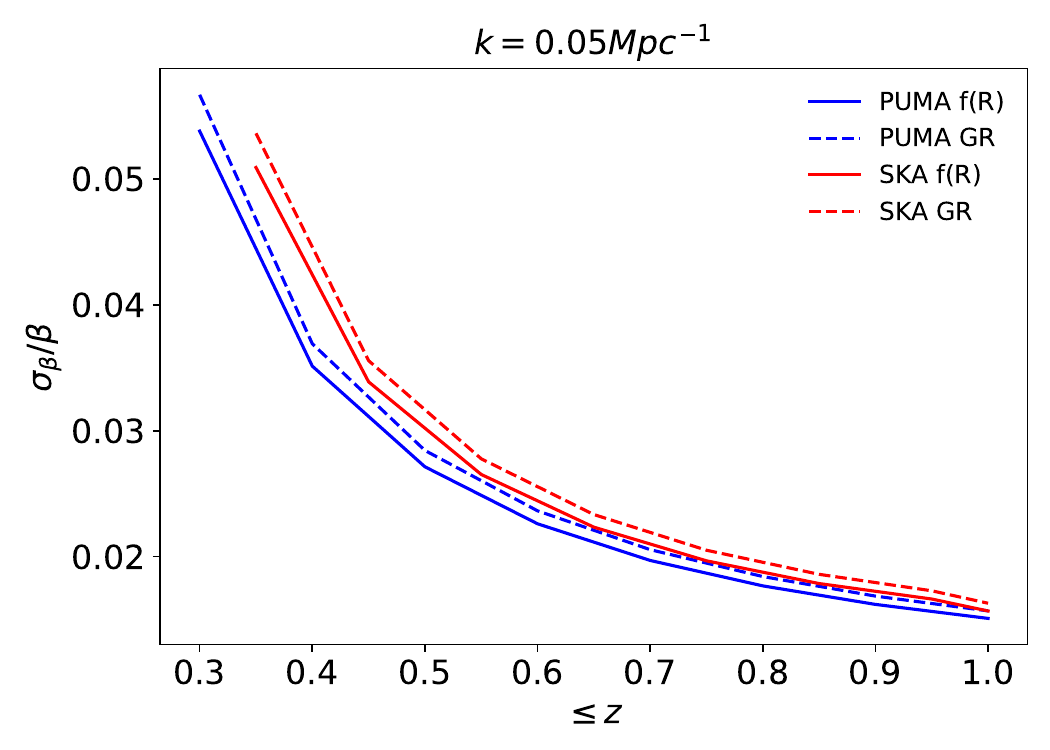}
    \caption{}
\end{subfigure}
\begin{subfigure}{0.33\textwidth}
    \centering
    \includegraphics[width=\linewidth]{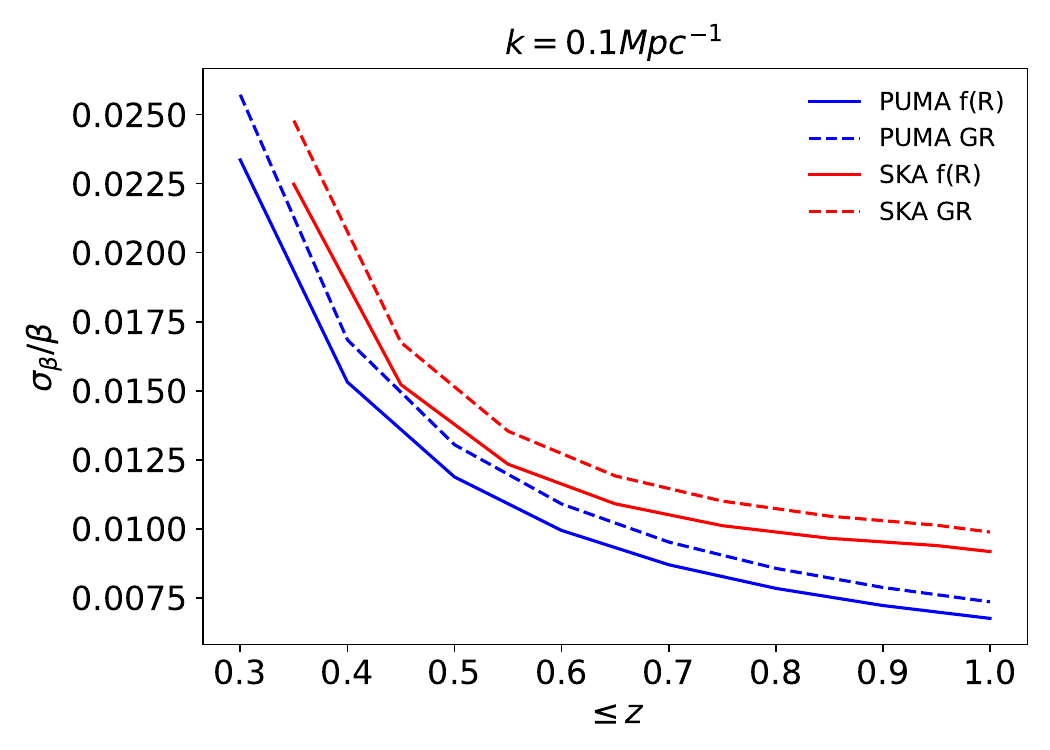}
    \caption{}
\end{subfigure}

\caption{Relative error of the full QS approximation across different Fourier scales. Panels (b,c,e,f) illustrate the $f(R)$ regime ($k = 0.05$ and $0.1\,\mathrm{Mpc}^{-1}$), where the relative error deviates significantly from the GR prediction. Panels (a,d) represent the GR-dominated regime, showing nearly identical relative errors and confirming theoretical consistency with survey expectations on these scales. These results remain consistent with the regime boundaries established in Fig.~(\ref{fig:Fr_GR_region}).}

\label{fig:full_relative_error}
\end{figure*}

\begin{figure*}
\centering

\begin{subfigure}{0.33\textwidth}
    \centering
    \includegraphics[width=\linewidth]{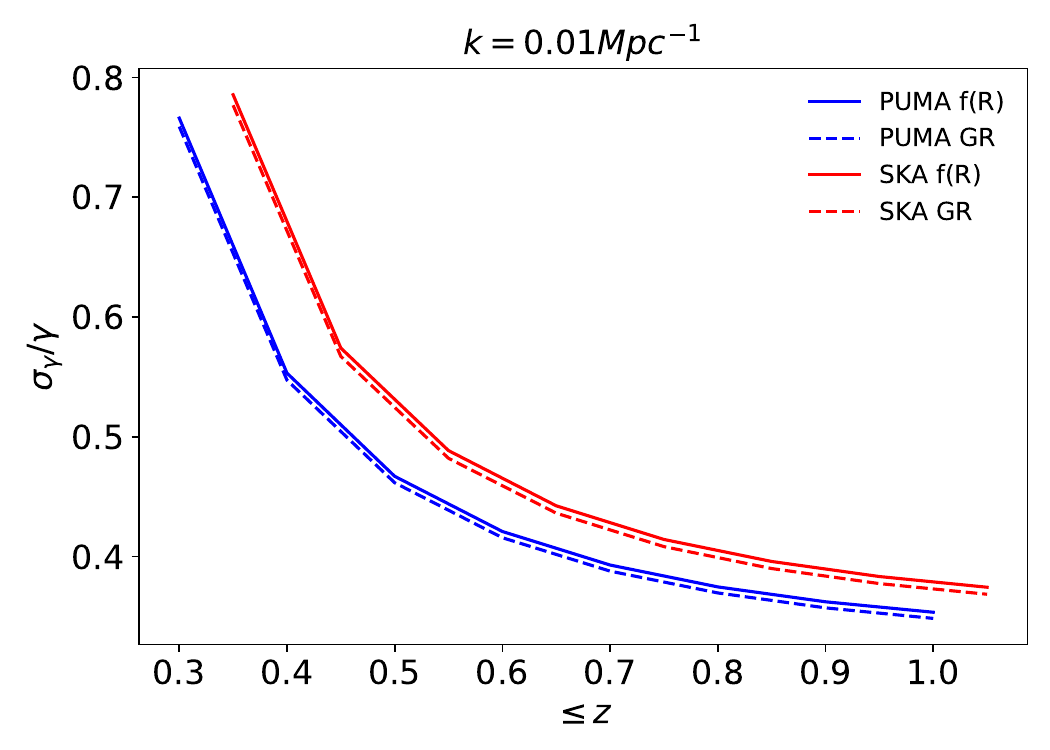} 
    \caption{}
\end{subfigure}
\begin{subfigure}{0.33\textwidth}
    \centering
    \includegraphics[width=\linewidth]{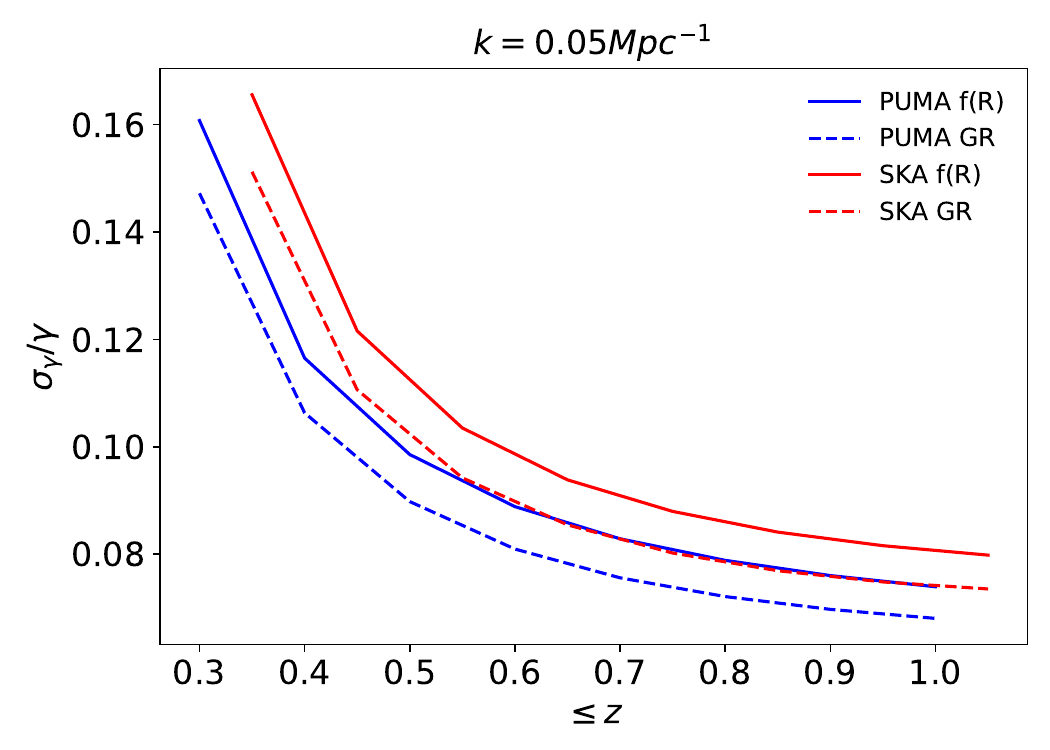}
    \caption{}
\end{subfigure}
\begin{subfigure}{0.33\textwidth}
    \centering
    \includegraphics[width=\linewidth]{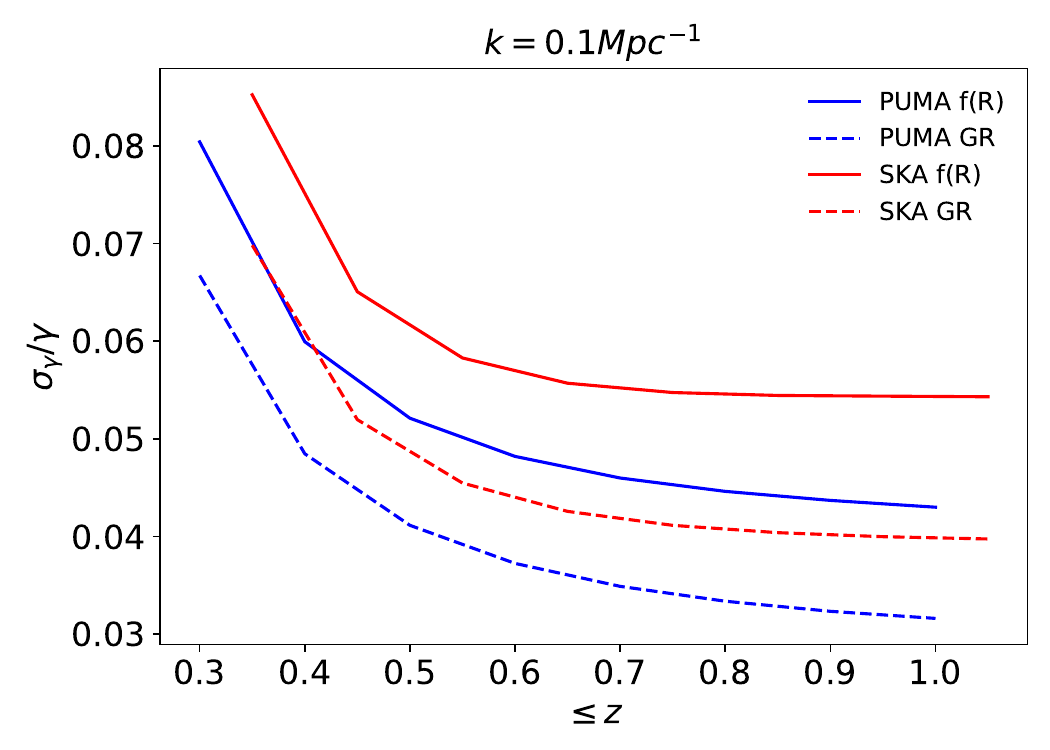}
    \caption{}
\end{subfigure}
\vspace{0.2cm}

\begin{subfigure}{0.33\textwidth}
    \centering
    \includegraphics[width=\linewidth]{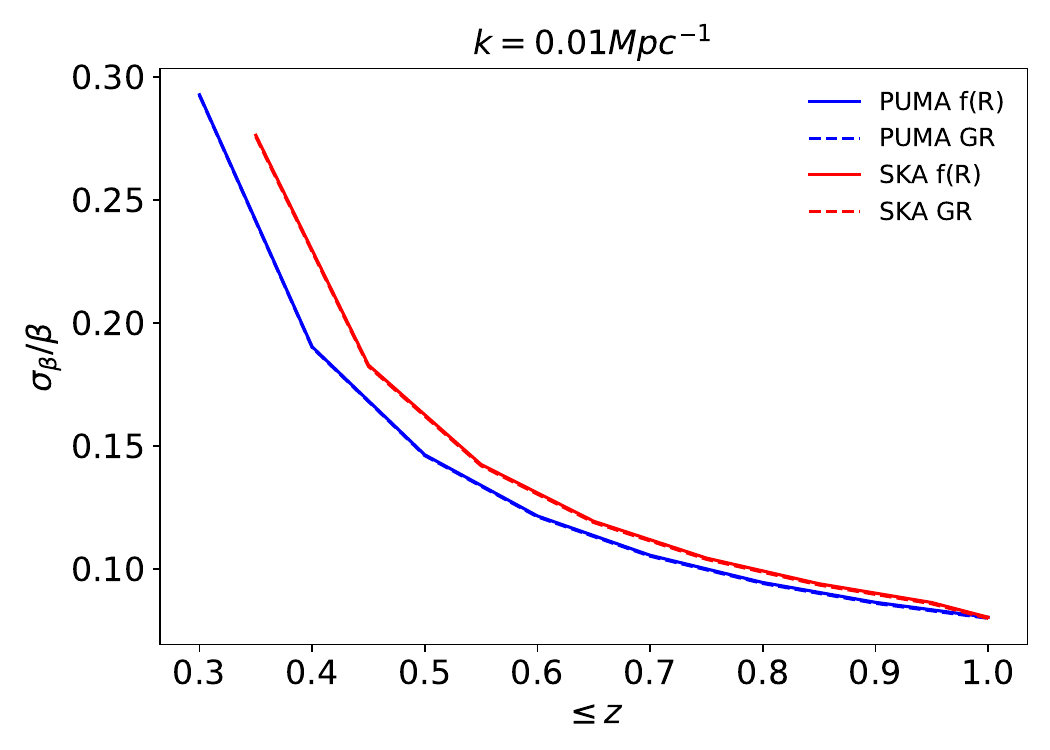}
    \caption{}
\end{subfigure}
\begin{subfigure}{0.33\textwidth}
    \centering
    \includegraphics[width=\linewidth]{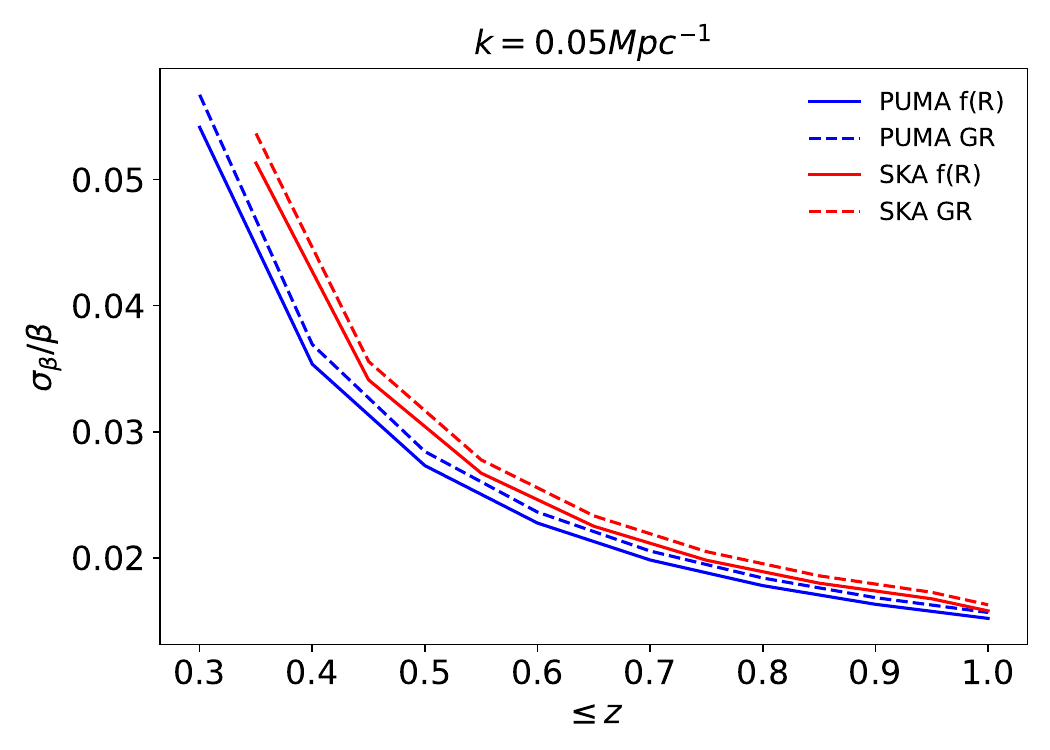}
    \caption{}
\end{subfigure}
\begin{subfigure}{0.33\textwidth}
    \centering
    \includegraphics[width=\linewidth]{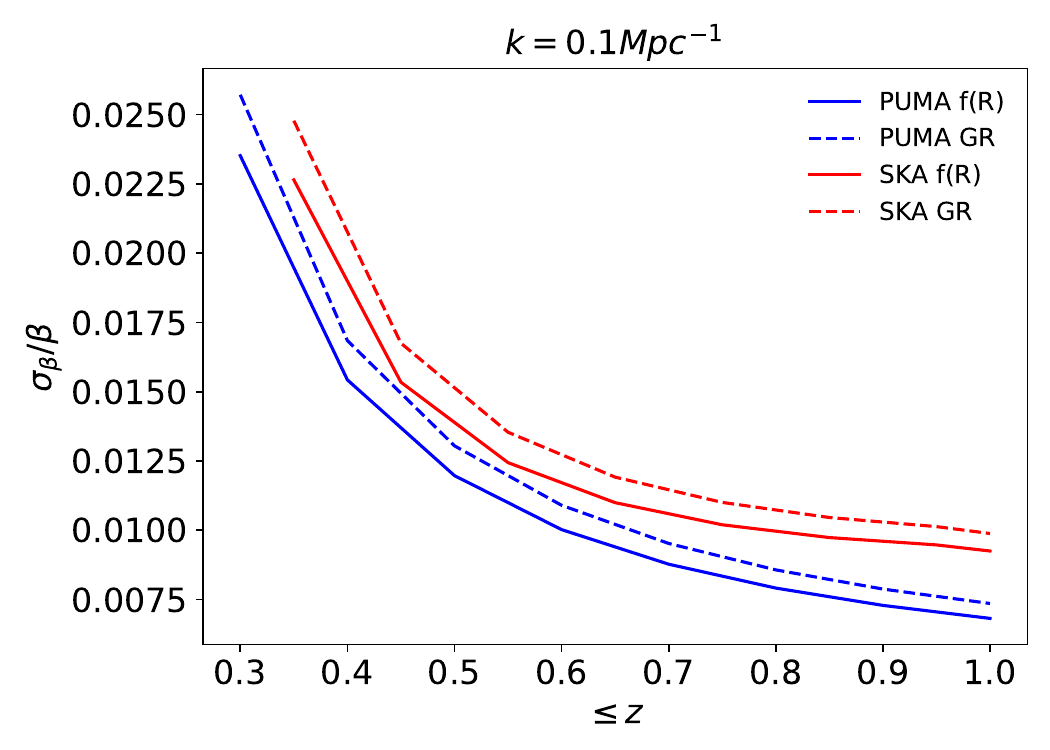}
    \caption{}
\end{subfigure}

\caption{Relative error of the Exact equation across different Fourier scales. Panels (b,c,e,f) illustrate the $f(R)$ regime ($k = 0.05$ and $0.1\,\mathrm{Mpc}^{-1}$), where the relative error deviates significantly from the GR prediction. Panels (a,d) represent the GR-dominated regime, showing nearly identical relative errors and confirming theoretical consistency with survey expectations on these scales. These results remain consistent with the regime boundaries established in Fig.~(\ref{fig:Fr_GR_region}).}
\label{fig:exact_relative_error}
\end{figure*}

In this section, we use Fisher formalism \citep{Long:2022dil, Karagiannis:2022ylq, amendola2024} to constrain $\gamma(k,z)$ and $\beta(k,z)$ parameters. 
The Fisher matrix for the parameter $\theta_i$ and $\theta_j$ is $F_{\theta_i\theta_j}(z)$ from power spectrum is

\begin{equation}\label{eq:Fisher_matrix}
\begin{aligned}
F_{\theta_i\theta_j}(z)=\frac{V_\mathrm{sur}}{16\pi^3}\int k^2\ d k\, d\mu\, d \phi\left(\frac{\partial P_{21}(\mathbf{k},z)}{\partial \theta_i}\right)\left(\frac{\partial P_{21}(\mathbf{k},z)}{\partial \theta_j}\right)\\ \frac{1}{(P_{21}(\mathbf{k},z))^2}\;.
\end{aligned}
\end{equation}
Here $V_\mathrm{sur}$ is the survey volume \citep{Long:2022dil, Volume:2024tvi}
\begin{equation}
V_\mathrm{sur}=\frac{4\pi}{3}f_\mathrm{sky}\left[\chi^3\left(z+\frac{\Delta z}{2}\right)-\chi^3\left(z-\frac{\Delta z}{2}\right)\right].
\end{equation}
$f_{\mathrm{sky}}$ is the sky fraction given in table \ref{table:Radio_arry}, $\chi(z)$ is the moving distance.
To constrain the growth-index parameters $\gamma(k,z)$ and $\beta(k,z)$ using Eq.~\eqref{eq:Fisher_matrix}, we utilise their scale dependence, which reduces the integration to a Dirac delta function in $k$. Due to the degeneracy between these parameters, we constrain them separately. The corresponding error on a parameter $\theta$ is given by $\sigma_{\theta} = \sqrt{(F^{-1})_{\theta\theta}}$. Assuming that the redshift bins are independent, we define the cumulative Fisher matrix as

\begin{equation}\label{eq:fisher_cumu}
F_{\theta_i\theta_j} = \sum_{z_i} F_{\theta_i\theta_j}(z_i)\;.   
\end{equation}
To ensure our results are applicable to upcoming radio surveys such as SKA and PUMA we restrict our analysis to comoving wave numbers in the range $0.001 \leq k \leq 0.1 \, \text{Mpc}^{-1}$. Furthermore, we adopt a specific initial condition, $x_{\text{in}}$, to guarantee that $\gamma > 0$ across all relevant modes. Finally, our calculations are confined to the low-redshift regime, where the observational signatures of $f(R)$ gravity become most prominent. In this paper, we consider the PUMA and SKA surveys, with HIRAX lying intermediate between them in observational scope. Using future 21 cm surveys and Fisher-matrix analysis \ref{eq:fisher_cumu}, we investigate both the exact model and the full QS model, and compare each with the $\Lambda$CDM model. As illustrated in Figs.~(\ref{fig:full_relative_error}) and (\ref{fig:exact_relative_error}), we estimate the relative errors on the two observable parameters, the growth index $\gamma(k,z)$ and the bias $\beta(k,z)$, for the exact and full QS cases from future 21 cm observations, and report the corresponding uncertainties. From the preceding Fisher analysis, we draw the following points:
\begin{enumerate}
    \item As demonstrated in Figs~\ref{fig:full_relative_error} and \ref{fig:exact_relative_error}, the standard $\Lambda$CDM model yields a smaller percentage relative error for the growth index $\gamma$ than both the full QS and exact $f(R)$ frameworks, independent of the mode. This is due to the evolutionary trajectory of $\gamma$ in $f(R)$ closely mimicking the $\Lambda$CDM case (see Fig.~\ref{fig:full growth}). While this leads to nearly identical error behaviours on large, linear scales  An ($k = 0.01\,\mathrm{Mpc}^{-1}$), a clear divergence between the $f(R)$ and GR error budgets becomes apparent on smaller scales ($k = 0.1\,\mathrm{Mpc}^{-1}$).
    
    \item In contrast, the relative error associated with the bias parameter $\beta$ yields tighter constraints in both the full quasi-static (QS) and exact $f(R)$ scenarios compared to the baseline $\Lambda$CDM model across all examined modes. Conversely, the relative error for the growth index $\gamma$ shows no observational improvement, remaining poorly constrained 
    relative to $\Lambda$CDM in both $f(R)$ frameworks across the same modes. 

Our results suggest that the parameter $\beta(k,z)$ exhibits sensitivity to $f(R)$ gravity, indicating that forthcoming 21-cm surveys could provide meaningful, though moderate, support for these modified gravity models. However, the corresponding constraints on $\gamma$ are currently limited, underscoring the necessity for further theoretical refinement. Future improvements in the modelling framework, combined with higher-precision observations, may enable $\gamma$ to serve as a more effective probe of modified gravity.

\end{enumerate}
   

   
\section{Conclusions}\label{sec:conclusion}

In this work, we investigated the observational signatures of $\Lambda$CDM-mimicking $f(R)$ gravity models using future 21\,cm intensity mapping surveys. Within this framework, the evolution of matter perturbations becomes both scale and redshift dependent, leading to a generalised growth index $\gamma(k,z)$ and growth-rate parameter $\beta(k,z)$ that differ from their standard General Relativistic (GR) counterparts. We explored these deviations by solving the perturbation equations in both the exact formulation and the full quasi-static (QS) approximation, and by analysing their observational consequences for future radio surveys such as SKA-Mid and PUMA. We first demonstrated that viable $\Lambda$CDM-mimicking $f(R)$ models can be constructed with suitable initial conditions that recover the GR limit at high redshift while generating measurable deviations at late times. In particular, the growth index exhibits a clear scale dependence over the observationally relevant range $10^{-2} \lesssim k \lesssim 10^{-1}\,\mathrm{Mpc}^{-1}$. The transition between the GR and modified-gravity regimes was shown 
to occur around $k \simeq 0.03\,\mathrm{Mpc}^{-1}$, beyond which the effects of modified gravity become increasingly significant. Using the Fisher matrix formalism applied to the 21\,cm power spectrum, we estimated the expected constraints on the generalised growth index $\gamma$ and the redshift-space distortion parameter $\beta$. Our results show that future 21\,cm surveys possess sufficient sensitivity to probe the scale-dependent growth predicted by $f(R)$ gravity. In 
particular, PUMA consistently provides tighter constraints than SKA-Mid owing to its larger survey volume and interferometric design. We also find that the relative uncertainties on $\gamma$ are generally larger in the modified-gravity scenario than in the GR limit, reflecting 
the increased parameter freedom introduced by scale-dependent growth. Conversely, the constraints on $\beta$ are comparatively tighter within the $f(R)$ framework. A comparison between the exact evolution equations and the full QS approximation shows that both approaches remain in close agreement over the scales relevant to upcoming 21\,cm surveys. The percentage difference between the two methods remains small across the redshift range considered, indicating that the QS approximation provides a reliable and computationally efficient description for observational 
forecasting in this class of models.

Overall, our analysis demonstrates that future 21\,cm intensity mapping experiments provide a promising avenue for testing deviations from GR on cosmological scales. Although the predicted deviations from the standard $\Lambda$CDM model are moderate, the scale dependence of the growth 
index offers a distinctive observational signature of $f(R)$ gravity. Future high-precision surveys combining 21\,cm observations with galaxy clustering and weak lensing measurements may therefore play an important role in distinguishing modified gravity models from standard cosmology.

\section*{Acknowledgements}
AS and BJ express their gratitude to the IIT Mandi for providing financial support and excellent computational facilities via the High Performance Computing (HPC) service. We would like to thank Dr. Rahul Kothari for useful discussions and comments on the manuscript. JW thanks the University of Cape Town for financial support. PKSD thanks the First Rand Bank (SA) for Financial Support.

\section*{Data Availability}
The theoretical framework and numerical code pipelines developed in this work are available via GitHub at \url{https://github.com/BhuUniverse/Modified-Gravity-using-21cm}. The underlying data sets generated during the current study are available from the corresponding author upon reasonable request.



\bibliographystyle{mnras}
\bibliography{example}

@article{Euclid:2025bxg,
    author = "Ocampo, I. and others",
    collaboration = "Euclid",
    title = "{Euclid: Forecasts on $\\Lambda$CDM consistency tests with growth rate data}",
    journal = "arXiv e-prints",
    eprint = "2507.22780",
    archivePrefix = "arXiv",
    primaryClass = "astro-ph.CO",
    year = "2025"
}

@article{Yin:2019rgm,
    author = "Yin, Zhao-Yu and Wei, Hao",
    title = "{Observational Constraints on Growth Index with Cosmography}",
    eprint = "1902.00289",
    archivePrefix = "arXiv",
    primaryClass = "astro-ph.CO",
    doi = "10.1140/epjc/s10052-019-7191-8",
    journal = "Eur. Phys. J. C",
    volume = "79",
    number = "8",
    pages = "698",
    year = "2019"
}

@article{Long:2022dil,
	archiveprefix = {arXiv},
	author = {Long, Heyang and Morales-Guti\'errez, Catalina and Montero-Camacho, Paulo and Hirata, Christopher M.},
	doi = {10.1093/mnras/stad2639},
	eprint = {2210.02385},
	journal = {Mon. Not. Roy. Astron. Soc.},
	number = {4},
	pages = {6036--6049},
	primaryclass = {astro-ph.CO},
	title = {{Impact of inhomogeneous reionization on post-reionization 21-cm intensity mapping measurement of cosmological parameters}},
	volume = {525},
	year = {2023}}

@article{Bharadwaj:2004nr,
	archiveprefix = {arXiv},
	author = {Bharadwaj, Somnath and Ali, Sheikh Saiyad},
	doi = {10.1111/j.1365-2966.2004.07907.x},
	eprint = {astro-ph/0401206},
	journal = {Mon. Not. Roy. Astron. Soc.},
	pages = {142},
	title = {{The CMBR fluctuations from HI perturbations prior to reionization}},
	volume = {352},
	year = {2004}}

@article{Karagiannis:2022ylq,
	archiveprefix = {arXiv},
	author = {Karagiannis, Dionysios and Maartens, Roy and Randrianjanahary, Liantsoa F.},
	doi = {10.1088/1475-7516/2022/11/003},
	eprint = {2206.07747},
	journal = {JCAP},
	pages = {003},
	primaryclass = {astro-ph.CO},
	title = {{Cosmological constraints from the power spectrum and bispectrum of 21cm intensity maps}},
	volume = {11},
	year = {2022}}

@article{Pritchard_2012,
	author = {Jonathan R Pritchard and Abraham Loeb},
	doi = {10.1088/0034-4885/75/8/086901},
	journal = {Reports on Progress in Physics},
	month = {jul},
	number = {8},
	pages = {086901},
	publisher = {IOP Publishing},
	title = {21 cm cosmology in the 21st century},
	url = {https://dx.doi.org/10.1088/0034-4885/75/8/086901},
	volume = {75},
	year = {2012}}

@article{Joshi:2025swr,
    author = "Joshi, Bhuwan and Kothari, Rahul",
    title = "{Constraining statistical isotropy using 21 cm power spectrum and bispectrum}",
    eprint = "2502.10717",
    archivePrefix = "arXiv",
    primaryClass = "astro-ph.CO",
    doi = "10.1088/1475-7516/2025/08/047",
    journal = "JCAP",
    volume = "08",
    pages = "047",
    year = "2025"
}

@article{Dossett:2010gq,
    author = "Dossett, Jason and Ishak, Mustapha and Moldenhauer, Jacob and Gong, Yungui and Wang, Anzhong and Gong, Yungui",
    title = "{Constraints on growth index parameters from current and future observations}",
    eprint = "1004.3086",
    archivePrefix = "arXiv",
    primaryClass = "astro-ph.CO",
    doi = "10.1088/1475-7516/2010/04/022",
    journal = "JCAP",
    volume = "04",
    pages = "022",
    year = "2010"
}

@book{21cm_BOOK,
	doi = {10.1088/2514-3433/ab4a73},
	editor = {Mesinger, Andrei},
	isbn = {978-0-7503-2236-2},
	publisher = {IOP Publishing},
	series = {2514-3433},
	title = {The Cosmic 21-cm Revolution},
	url = {https://dx.doi.org/10.1088/2514-3433/ab4a73},
	year = {2019}}

@article{SKA,
	author = {Bacon, David J. and Battye, Richard A. and Bull, Philip and Camera, Stefano and Ferreira, Pedro G. and Harrison, Ian and Parkinson, David and Pourtsidou, Alkistis and Santos, M{\'a}rio G. and Wolz, Laura and Abdalla, Filipe and Akrami, Yashar and Alonso, David and Andrianomena, Sambatra and Ballardini, Mario and Bernal, Jos{\'e} Luis and Bertacca, Daniele and Bengaly, Carlos A. P. and Bonaldi, Anna and Bonvin, Camille and Brown, Michael L. and Chapman, Emma and Chen, Song and Chen, Xuelei and Cunnington, Steven and Davis, Tamara M. and Dickinson, Clive and Fonseca, Jos{\'e} and Grainge, Keith and Harper, Stuart and Jarvis, Matt J. and Maartens, Roy and Maddox, Natasha and Padmanabhan, Hamsa and Pritchard, Jonathan R. and Raccanelli, Alvise and Rivi, Marzia and Roychowdhury, Sambit and Sahl{\'e}n, Martin and Schwarz, Dominik J. and Siewert, Thilo M. and Viel, Matteo and Villaescusa-Navarro, Francisco and Xu, Yidong and Yamauchi, Daisuke and Zuntz, Joe},
	doi = {10.1017/pasa.2019.51},
	issn = {1448-6083},
	journal = {Publications of the Astronomical Society of Australia},
	publisher = {Cambridge University Press (CUP)},
	title = {Cosmology with Phase 1 of the Square Kilometre Array Red Book 2018: Technical specifications and performance forecasts},
	url = {http://dx.doi.org/10.1017/pasa.2019.51},
	volume = {37},
	year = {2020}}

@article{21cm_review,
	archiveprefix = {arXiv},
	author = {Furlanetto, Steven and Oh, S. Peng and Briggs, Frank},
	doi = {10.1016/j.physrep.2006.08.002},
	eprint = {astro-ph/0608032},
	journal = {Phys. Rept.},
	pages = {181--301},
	title = {{Cosmology at Low Frequencies: The 21 cm Transition and the High-Redshift Universe}},
	volume = {433},
	year = {2006}}

@misc{amendola2024,
	author = {Luca Amendola},
	note = {University of Heidelberg},
	title = {Statistical Methods Lecture Notes},
	url = {https://www.thphys.uni-heidelberg.de/~amendola/teaching/compstat-hd.pdf},
	urldate = {2024-05-23},
	year = {2024}}

@article{CosmicVisions21cm:2018rfq,
    author = "Ansari, R{\'e}za and others",
    collaboration = "Cosmic Visions 21 cm",
    title = "{Inflation and Early Dark Energy with a Stage II Hydrogen Intensity Mapping experiment}",
    journal = "arXiv e-prints",
    eprint = "1810.09572",
    archivePrefix = "arXiv",
    primaryClass = "astro-ph.CO",
    year = "2018"
}

@article{Bull:2014rha,
    author = "Bull, Philip and Ferreira, Pedro G. and Patel, Prina and Santos, Mario G.",
    title = "{Late-time cosmology with 21cm intensity mapping experiments}",
    eprint = "1405.1452",
    archivePrefix = "arXiv",
    primaryClass = "astro-ph.CO",
    doi = "10.1088/0004-637X/803/1/21",
    journal = "Astrophys. J.",
    volume = "803",
    number = "1",
    pages = "21",
    year = "2015"
}

@article{SKA:2018ckk,
    author = "Bacon, David J. and others",
    collaboration = "SKA",
    title = "{Cosmology with Phase 1 of the Square Kilometre Array: Red Book 2018: Technical specifications and performance forecasts}",
    eprint = "1811.02743",
    archivePrefix = "arXiv",
    primaryClass = "astro-ph.CO",
    doi = "10.1017/pasa.2019.51",
    journal = "Publ. Astron. Soc. Austral.",
    volume = "37",
    pages = "e007",
    year = "2020"
}

@article{Gong:2011qf,
    author = "Gong, Yan and Chen, Xuelei and Silva, Marta and Cooray, Asantha and Santos, Mario G.",
    title = "{The OH line contamination of 21 cm intensity fluctuation measurements for z=1\textasciitilde{}4}",
    eprint = "1108.0947",
    archivePrefix = "arXiv",
    primaryClass = "astro-ph.CO",
    doi = "10.1088/2041-8205/740/1/L20",
    journal = "Astrophys. J. Lett.",
    volume = "740",
    pages = "L20",
    year = "2011"
}

@article{Santos:2015gra,
    author = "Santos, Mario G. and others",
    editor = "Bourke, Tyler L. and others",
    title = "{Cosmology from a SKA HI intensity mapping survey}",
    eprint = "1501.03989",
    archivePrefix = "arXiv",
    primaryClass = "astro-ph.CO",
    doi = "10.22323/1.215.0019",
    journal = "PoS",
    volume = "AASKA14",
    pages = "019",
    year = "2015"
}

@article{Volume:2024tvi,
    author = "Rossiter, Samantha J. and Camera, Stefano and Clarkson, Chris and Maartens, Roy",
    title = "{Decoupling local primordial non-Gaussianity from relativistic effects in the galaxy bispectrum}",
    eprint = "2407.06301",
    archivePrefix = "arXiv",
    primaryClass = "astro-ph.CO",
    doi = "10.1088/1475-7516/2025/07/055",
    journal = "JCAP",
    volume = "07",
    pages = "055",
    year = "2025"
}

@article{Hall:2012wd,
    author = "Hall, Alex and Bonvin, Camille and Challinor, Anthony",
    title = "{Testing General Relativity with 21-cm intensity mapping}",
    eprint = "1212.0728",
    archivePrefix = "arXiv",
    primaryClass = "astro-ph.CO",
    doi = "10.1103/PhysRevD.87.064026",
    journal = "Phys. Rev. D",
    volume = "87",
    number = "6",
    pages = "064026",
    year = "2013"
}

@article{FoG:1996cd,
    author = "Ballinger, W. E. and Peacock, J. A. and Heavens, A. F.",
    title = "{Measuring the cosmological constant with redshift surveys}",
    eprint = "astro-ph/9605017",
    archivePrefix = "arXiv",
    doi = "10.1093/mnras/282.3.877",
    journal = "Mon. Not. Roy. Astron. Soc.",
    volume = "282",
    pages = "877--888",
    year = "1996"
}

@article{Darkmatter_21cm:2023slb,
    author = "Facchinetti, Ga\'etan and Lopez-Honorez, Laura and Qin, Yuxiang and Mesinger, Andrei",
    title = "{21cm signal sensitivity to dark matter decay}",
    eprint = "2308.16656",
    archivePrefix = "arXiv",
    primaryClass = "astro-ph.CO",
    doi = "10.1088/1475-7516/2024/01/005",
    journal = "JCAP",
    volume = "01",
    pages = "005",
    year = "2024"
}

@article{Darkmatter_21cm_2:2016sur,
    author = "Lopez-Honorez, Laura and Mena, Olga and Molin\'e, \'Angeles and Palomares-Ruiz, Sergio and Vincent, Aaron C.",
    title = "{The 21 cm signal and the interplay between dark matter annihilations and astrophysical processes}",
    eprint = "1603.06795",
    archivePrefix = "arXiv",
    primaryClass = "astro-ph.CO",
    reportNumber = "CFTP-16-007, IFIC-16-16, IPPP-16-20",
    doi = "10.1088/1475-7516/2016/08/004",
    journal = "JCAP",
    volume = "08",
    pages = "004",
    year = "2016"
}

@article{Ali-Haimoud:2013hpa,
    author = {Ali-Ha\"\i{}moud, Yacine and Meerburg, P. Daniel and Yuan, Sihan},
    title = "{New light on 21 cm intensity fluctuations from the dark ages}",
    eprint = "1312.4948",
    archivePrefix = "arXiv",
    primaryClass = "astro-ph.CO",
    doi = "10.1103/PhysRevD.89.083506",
    journal = "Phys. Rev. D",
    volume = "89",
    number = "8",
    pages = "083506",
    year = "2014"
}

@article{Mondal:2023xjx,
    author = "Mondal, Rajesh and Barkana, Rennan",
    title = "{Prospects for precision cosmology with the 21\,cm signal from the dark ages}",
    eprint = "2305.08593",
    archivePrefix = "arXiv",
    primaryClass = "astro-ph.CO",
    doi = "10.1038/s41550-023-02057-y",
    journal = "Nature Astron.",
    volume = "7",
    number = "9",
    pages = "1025--1030",
    year = "2023"
}

@article{Saurabh_RRI:2021mxo,
    author = "Singh, Saurabh and Nambissan T., Jishnu and Subrahmanyan, Ravi and Udaya Shankar, N. and Girish, B. S. and Raghunathan, A. and Somashekar, R. and Srivani, K. S. and Sathyanarayana Rao, Mayuri",
    title = "{On the detection of a cosmic dawn signal in the radio background}",
    eprint = "2112.06778",
    archivePrefix = "arXiv",
    primaryClass = "astro-ph.CO",
    doi = "10.1038/s41550-022-01610-5",
    journal = "Nature Astron.",
    volume = "6",
    number = "5",
    pages = "607--617",
    year = "2022"
}

@article{Ankita_Bera:2022vhw,
    author = "Bera, Ankita and Ghara, Raghunath and Chatterjee, Atrideb and Datta, Kanan K. and Samui, Saumyadip",
    title = "{Studying cosmic dawn using redshifted HI 21-cm signal: A brief review}",
    eprint = "2210.12164",
    archivePrefix = "arXiv",
    primaryClass = "astro-ph.CO",
    doi = "10.1007/s12036-022-09904-w",
    journal = "J. Astrophys. Astron.",
    volume = "44",
    number = "1",
    pages = "10",
    year = "2023"
}

@article{Santos:2015noise,
    author = "Santos, Mario G. and others",
    editor = "Bourke, Tyler L. and others",
    title = "{Cosmology from a SKA HI intensity mapping survey}",
    eprint = "1501.03989",
    archivePrefix = "arXiv",
    primaryClass = "astro-ph.CO",
    doi = "10.22323/1.215.0019",
    journal = "PoS",
    volume = "AASKA14",
    pages = "019",
    year = "2015"
}

@article{Jolicoeur:2020eup,
    author = "Jolicoeur, Sheean and Maartens, Roy and De Weerd, Eline M. and Umeh, Obinna and Clarkson, Chris and Camera, Stefano",
    title = "{Detecting the relativistic bispectrum in 21cm intensity maps}",
    eprint = "2009.06197",
    archivePrefix = "arXiv",
    primaryClass = "astro-ph.CO",
    doi = "10.1088/1475-7516/2021/06/039",
    journal = "JCAP",
    volume = "06",
    pages = "039",
    year = "2021"
}

@article{Chakraborty:2021jku,
    author = "Chakraborty, Saikat and MacDevette, Kelly and Dunsby, Peter",
    title = "{A model independent approach to the study of $f(R)$ cosmologies with expansion histories close to $\Lambda$CDM}",
    eprint = "2103.02274",
    archivePrefix = "arXiv",
    primaryClass = "gr-qc",
    doi = "10.1103/PhysRevD.103.124040",
    journal = "Phys. Rev. D",
    volume = "103",
    number = "12",
    pages = "124040",
    year = "2021"
}

@article{MacDevette:2024wpg,
    author = "MacDevette, Kelly and Worsley, Jess and Dunsby, Peter and Chakraborty, Saikat",
    title = "{A model-independent approach to the study of structure growth in f(R) gravity}",
    eprint = "2408.03998",
    archivePrefix = "arXiv",
    primaryClass = "gr-qc",
    doi = "10.1093/mnras/staf168",
    journal = "Mon. Not. Roy. Astron. Soc.",
    volume = "537",
    number = "3",
    pages = "2471--2495",
    year = "2025"
}

@article{Chakraborty:2025lkz,
    author = "Chakraborty, Saikat and Burikham, Piyabut",
    title = "{Theory space and stability analysis of General Relativistic cosmological solutions in modified gravity}",
    eprint = "2509.15762",
    archivePrefix = "arXiv",
    primaryClass = "gr-qc",
    doi = "10.1140/epjc/s10052-026-15390-z",
    journal = "Eur. Phys. J. C",
    volume = "86",
    number = "3",
    pages = "280",
    year = "2026"
}

@article{Chakraborty:2022evc,
    author = "Chakraborty, Saikat and Gregoris, Daniele and Mishra, B.",
    title = "{On the uniqueness of {\ensuremath{\Lambda}}CDM-like evolution for homogeneous and isotropic cosmology in General Relativity}",
    eprint = "2208.04596",
    archivePrefix = "arXiv",
    primaryClass = "gr-qc",
    doi = "10.1016/j.physletb.2023.137962",
    journal = "Phys. Lett. B",
    volume = "842",
    pages = "137962",
    year = "2023"
}

@ARTICLE{Ma1995,
       author = {{Ma}, Chung-Pei and {Bertschinger}, Edmund},
        title = "{Cosmological Perturbation Theory in the Synchronous and Conformal Newtonian Gauges}",
      journal = {\apj},
         year = 1995,
        month = dec,
       volume = {455},
        pages = {7},
          doi = {10.1086/176550},
archivePrefix = {arXiv},
       eprint = {astro-ph/9506072},
 primaryClass = {astro-ph},
       adsurl = {https://ui.adsabs.harvard.edu/abs/1995ApJ...455....7M}
}

@article{10.1093/mnras/150.1.1,
    author = {Buchdahl, H. A.},
    title = {Non-Linear Lagrangians and Cosmological Theory},
    journal = {Monthly Notices of the Royal Astronomical Society},
    volume = {150},
    number = {1},
    pages = {1-8},
    year = {1970},
    month = {09},
    issn = {0035-8711},
    doi = {10.1093/mnras/150.1.1},
    url = {https://doi.org/10.1093/mnras/150.1.1},
    eprint = {https://academic.oup.com/mnras/article-pdf/150/1/1/8075909/mnras150-0001.pdf},
}

@book{eddington1923,
  author    = {Eddington, A. S.},
  title     = {The Mathematical Theory of Relativity},
  publisher = {Cambridge University Press},
  year      = {1923},
  address   = {Cambridge}
}

@article{Starobinsky:1980te,
    author = "Starobinsky, Alexei A.",
    editor = "Khalatnikov, I. M. and Mineev, V. P.",
    title = "{A New Type of Isotropic Cosmological Models Without Singularity}",
    doi = "10.1016/0370-2693(80)90670-X",
    journal = "Phys. Lett. B",
    volume = "91",
    pages = "99--102",
    year = "1980"
}

@article{PhysRevD.76.064004,
  title = {Models of $f(R)$ cosmic acceleration that evade solar system tests},
  author = {Hu, Wayne and Sawicki, Ignacy},
  journal = {Phys. Rev. D},
  volume = {76},
  issue = {6},
  pages = {064004},
  numpages = {13},
  year = {2007},
  month = {Sep},
  publisher = {American Physical Society},
  doi = {10.1103/PhysRevD.76.064004},
  url = {https://link.aps.org/doi/10.1103/PhysRevD.76.064004}
}

@article{PhysRevD.16.953,
  title = {Renormalization of higher-derivative quantum gravity},
  author = {Stelle, K. S.},
  journal = {Phys. Rev. D},
  volume = {16},
  issue = {4},
  pages = {953--969},
  numpages = {0},
  year = {1977},
  month = {Aug},
  publisher = {American Physical Society},
  doi = {10.1103/PhysRevD.16.953},
  url = {https://link.aps.org/doi/10.1103/PhysRevD.16.953}
}

@article{PhysRevD.70.043528,
  title = {Is cosmic speed-up due to new gravitational physics?},
  author = {Carroll, Sean M. and Duvvuri, Vikram and Trodden, Mark and Turner, Michael S.},
  journal = {Phys. Rev. D},
  volume = {70},
  issue = {4},
  pages = {043528},
  numpages = {5},
  year = {2004},
  month = {Aug},
  publisher = {American Physical Society},
  doi = {10.1103/PhysRevD.70.043528},
  url = {https://link.aps.org/doi/10.1103/PhysRevD.70.043528}
}

@article{Carloni_2005,
   title={Cosmological dynamics of
                    
                      R
                      n
                    
                    gravity,	arXiv:gr-qc/0410046},
   volume={22},
   ISSN={1361-6382},
   url={http://dx.doi.org/10.1088/0264-9381/22/22/011},
   DOI={10.1088/0264-9381/22/22/011},
   number={22},
   journal={Classical and Quantum Gravity},
   publisher={IOP Publishing},
   author={Carloni, S and Dunsby, P K S and Capozziello, S and Troisi, A},
   year={2005},
   month=oct, pages={4839–4868} }

@article{article,
author = {Chakraborty, Saikat and MacDevette, Kelly and Dunsby, Peter},
year = {2021},
month = {06},
pages = {},
title = {Model independent approach to the study of f ( R ) cosmologies with expansion histories close to Λ CDM},
volume = {103},
journal = {Physical Review D},
doi = {10.1103/PhysRevD.103.124040}
}

@article{Bull2015_21cm,
    author = {Bull, Philip and Ferreira, Pedro G. and Patel, Prina and Santos, Mario G.},
    title = {Late-time cosmology with 21cm intensity mapping experiments},
    eprint = {1405.1452},
    archivePrefix = {arXiv},
    primaryClass = {astro-ph.CO},
    doi = {10.1088/0004-637X/803/1/21},
    journal = {Astrophys. J.},
    volume = {803},
    number = {1},
    pages = {21},
    year = {2015}
}

@article{Karagiannis:2024noise,
    author = "Karagiannis, Dionysios and Maartens, Roy and Saito, Shun and Fonseca, Jos\'e and Camera, Stefano and Clarkson, Chris",
    title = "{Squeezing information from radio surveys to probe the primordial Universe}",
    journal = "arXiv e-prints",
    eprint = "2406.00117",
    archivePrefix = "arXiv",
    primaryClass = "astro-ph.CO",
    year = "2024"
}

@article{10.1093/mnras/227.1.1,
    author = {Kaiser, Nick},
    title = {Clustering in real space and in redshift space},
    journal = {Monthly Notices of the Royal Astronomical Society},
    volume = {227},
    number = {1},
    pages = {1-21},
    year = {1987},
    month = {07},
    issn = {0035-8711},
    doi = {10.1093/mnras/227.1.1},
    url = {https://doi.org/10.1093/mnras/227.1.1},
    eprint = {https://academic.oup.com/mnras/article-pdf/227/1/1/18522208/mnras227-0001.pdf},
}

@article{Dicke1961,
  author  = {Dicke, R. H.},
  title   = {Dirac's Cosmology and Mach's Principle},
  journal = {Nature},
  volume  = {192},
  pages   = {440--441},
  year    = {1961},
  doi     = {10.1038/192440a0}
}

@inproceedings{Steinhardt1998,
  author    = {Steinhardt, P. J.},
  title     = {Cosmological Challenges for the 21st Century},
  booktitle = {Critical Problems in Physics},
  editor    = {Fitch, V. L. and Marlow, D. R.},
  publisher = {Princeton University Press},
  year      = {1997},
  note      = {Often cited in late 1990s literature regarding the coincidence problem prior to the formal introduction of quintessence.}
}

@article{Adler1995,
  author  = {Adler, R. J. and Casey, B. and Jacob, O. C.},
  title   = {Vacuum catastrophe: An elementary exposition of the cosmological constant problem},
  journal = {American Journal of Physics},
  volume  = {63},
  number  = {7},
  pages   = {620--626},
  year    = {1995},
  doi     = {10.1119/1.17850}
}

@article{Carloni2024,
  author        = {Carloni, Y. and Luongo, O. and Muccino, M.},
  title         = {Does dark energy really revive using DESI 2024 data?},
  journal       = {arXiv e-prints},
  year          = {2024},
  eprint        = {2404.12068},
  archivePrefix = {arXiv},
  primaryClass  = {gr-qc}
}

@article{Luongo2024,
  author        = {Luongo, O. and Muccino, M.},
  title         = {Relaxing the $H_0$ tension through DESI 2024 data},
  journal       = {arXiv e-prints},
  year          = {2024},
  note          = {Representative citation for the Luongo & Muccino 2024 constraints.}
}

@article{Tada2024,
  author  = {Tada, Y. and Terada, T.},
  title   = {Quintessential interpretation of the evolving dark energy in light of DESI},
  journal = {Physical Review D},
  volume  = {109},
  number  = {12},
  pages   = {L121305},
  year    = {2024},
  doi     = {10.1103/PhysRevD.109.L121305}
}

@article{Carloni:2006mr,
    author = "Carloni, Sante and Dunsby, Peter K. S.",
    editor = "Sola, Joan",
    title = "{A Dynamical system approach to higher order gravity}",
    eprint = "gr-qc/0611122",
    archivePrefix = "arXiv",
    doi = "10.1088/1751-8113/40/25/S40",
    journal = "J. Phys. A",
    volume = "40",
    pages = "6919--6926",
    year = "2007"
}

@article{Carloni:2007yv,
    author = "Carloni, S. and Dunsby, P. K. S. and Troisi, A.",
    title = "{The Evolution of density perturbations in f(R) gravity}",
    eprint = "0707.0106",
    archivePrefix = "arXiv",
    primaryClass = "gr-qc",
    doi = "10.1103/PhysRevD.77.024024",
    journal = "Phys. Rev. D",
    volume = "77",
    pages = "024024",
    year = "2008"
}

@article{Ananda:2008tx,
    author = "Ananda, KishoreN. and Carloni, Sante and Dunsby, Peter K. S.",
    title = "{A detailed analysis of structure growth in $f(R)$ theories of gravity}",
    eprint = "0809.3673",
    archivePrefix = "arXiv",
    primaryClass = "astro-ph",
    doi = "10.1088/0264-9381/26/23/235018",
    journal = "Class. Quant. Grav.",
    volume = "26",
    pages = "235018",
    year = "2009"
}






\bsp	
\label{lastpage}
\end{document}